\documentclass[12pt]{scrartcl}

\usepackage{hyperref}
\usepackage{amssymb}
\usepackage{amsmath}
\usepackage{amsthm}
\usepackage{bbm}
\usepackage{graphicx,subfigure}
\usepackage{float}
\usepackage{units}
\usepackage{algorithm,algorithmic}
\usepackage{stmaryrd}
\usepackage{bm}

\newcommand{\ignore}[1]{}
\newcommand{\tit}[1]{\text{\itshape{#1}}}
\newcommand{\shift}{\tit{shift}\,}
\newcommand{\rightpos}{\tit{right}\,}
\newcommand{\daaval}{\tit{value}}

\newcommand{\N}{\mathbbmss{N}}
\newcommand{\Z}{\mathbbmss{Z}}
\newcommand{\R}{\mathbbmss{R}}
\newcommand{\OO}{\mathcal{O}}
\newcommand{\prob}{\mathbbmss{P}}

\newcommand{\dist}{\mathcal{L}}
\newcommand{\mc}[1]{\mathcal{#1}}
 
\newcommand{\given}{\,|\,}
\newcommand{\mboxspace}[1]{\mbox{\hspace{1em} #1 \hspace{1em}}}
\newcommand{\paa}{\mc{P}}
\newcommand{\daa}{\mc{D}}
\newcommand{\ps}{\mc{PS}}
\newcommand{\len}[1]{{\vert {#1} \vert}}
\newcommand{\emptystring}{\varepsilon}
\newcommand{\substr}[3]{{#1}[{#2}\ldots{#3}]}
\newcommand{\chr}[2]{#1[#2]}
\newcommand{\suffix}[2]{#1[#2..]}
\newcommand{\prefix}[2]{#1[..#2]}
\newcommand{\iverl}{\llbracket}
\newcommand{\iverr}{\rrbracket}
\newcommand{\dirac}{\bm\delta}
\newcommand{\ie}{i.e.\ }
\newcommand{\iid}{i.i.d.\ }

\newcommand{\stateset}{\mathcal{Q}}
\newcommand{\emiset}{\mathcal{E}}
\newcommand{\valset}{\mathcal{V}}
\newcommand{\valsize}{\vartheta}
\newcommand{\stateproc}{Q}
\newcommand{\emiproc}{E}
\newcommand{\valproc}{V}
\newcommand{\state}{q}
\newcommand{\emi}{e}
\newcommand{\val}{v}
\newcommand{\op}{\theta}

\newcommand{\emidist}{\mu}       
\newcommand{\paatrans}{T}        
\newcommand{\daaemi}{\eta}       
\newcommand{\deltajoint}{\bar{\delta}}  
\newcommand{\totalsteps}{n}      
\newcommand{\occcount}{k}        
\newcommand{\pattern}{p}         
\newcommand{\patlen}{m}          
\newcommand{\tmax}{\totalsteps}  

\newcommand{\paatuple}{\big(\stateset,\state_0,\paatrans,\valset,\val_0, \emiset, \emidist=(\emidist_\state)_{\state\in\stateset},\op=(\op_\state)_{\state\in\stateset}\big)}
\newcommand{\paatuplebrief}{(\stateset,\state_0,\paatrans,\valset,\val_0, \emiset, \emidist,\op)}

\newcommand{\windowcost}[1]{\xi^{#1}_{\text{\textvisiblespace}}}
\newcommand{\maxwindowcost}[1]{\hat{\xi}^{#1}_{\text{\textvisiblespace}}}
\newcommand{\cost}[1]{\xi^{#1}}

\newcommand{\fragmentX}{F_{\star}}
\newcommand{\fragmentXdist}{\nu_{\star}}

\newtheorem{theorem}{Theorem}[section]
\newtheorem{lemma}[theorem]{Lemma}
\newtheorem{example}[theorem]{Example}
\newtheorem{problem}[theorem]{Problem}
\newtheorem{definition}[theorem]{Definition}
\newtheorem{remark}[theorem]{Remark}

\floatstyle{ruled}  
\newfloat{recurrence}{ht}{test}[section]
\floatname{recurrence}{\textbf{Recurrence}}

\title{Probabilistic Arithmetic Automata and their Applications\thanks{Parts of this article have been published in conference proceedings~\cite{DBLP:conf/recomb/KaltenbachBR06,Marschall2008,Herms2008,Marschall2009,Marschall2010}. An extended version of one of these articles \cite{Marschall2010} has been submitted to a journal. A preprint is available on arXiv \cite{Marschall2010b}.}}

\author{Tobias Marschall\thanks{Bioinformatics for High-Throughput Technologies, Computer Science 11, TU Dortmund, 44221~Dortmund, Germany} 
  \and Inke Herms\thanks{Genome Informatics, Faculty of Technology, Bielefeld University, 33501~Bielefeld, Germany} 
  \and Hans-Michael Kaltenbach\thanks{Department of Biosystems Science and Engineering, ETH Zurich, 4058~Basel, Switzerland} 
  \and Sven Rahmann${}^\dagger$
}

\begin{document}

\maketitle

\begin{abstract}
\textbf{Abstract.} We present \emph{probabilistic arithmetic automata (PAAs)}, a general model to describe chains of operations whose operands depend on chance, along with two different algorithms to exactly calculate the distribution of the results obtained by such probabilistic calculations. 
PAAs provide a unifying framework to approach many problems arising in computational biology and elsewhere. Here, we present five different applications, namely (1)~pattern matching statistics on random texts, including the computation of the distribution of occurrence counts, waiting time and clump size under HMM background models; 
(2)~exact analysis of window-based pattern matching algorithms; (3)~sensitivity of filtration seeds used to detect candidate sequence alignments; (4)~length and mass statistics of peptide fragments resulting from enzymatic cleavage reactions; and (5)~read length statistics of 454~sequencing reads. The diversity of these applications indicates the flexibility and unifying character of the presented framework.

While the construction of a PAA depends on the particular application, we single out a frequently applicable construction method for pattern statistics: We introduce \emph{deterministic arithmetic automata (DAAs)} to model deterministic calculations on sequences, and demonstrate how to construct a PAA from a given DAA and a finite-memory random text model. We show how to transform a finite automaton into a DAA and then into the corresponding PAA.

\end{abstract}


\section{Introduction}\label{sec:introduction}
In many applications, processes can be modeled as chains of operations working on operands that are drawn probabilistically. 
As an example, let us consider a simple dice game. 
Suppose you have a bag containing three dice, a 6-faced, a 12-faced, and a 20-faced die. 
Now a die is drawn from the bag, rolled, and put back.
This procedure is repeated $\totalsteps$~times. 
In the end one may, for example, be interested in the distribution of the maximum number observed. Many variants can be thought of, for instance, we might start with a value of~0 and each die might be associated with an operation, e.g. the spots seen on the 6-faced die might be subtracted from the current value and the spots on the 12-faced and 20-faced dice might be added. 
In addition to the distribution of values after $\totalsteps$~rolls, we can ask for the distribution of the waiting time for reaching a value above a given threshold. 

The goal of this article is to establish a general formal framework, referred to as  \emph{probabilistic arithmetic automata (PAAs)}, to directly model such systems and answer the posed questions. 
We emphasize that we are not interested in simulation studies or approximations to these distributions, but in an exact computation up to machine accuracy.
Further, we show that problems from diverse applications, especially from computational biology, can be conveniently solved with PAAs in a unified way, whereas they are so far treated heterogeneously in the literature.
This article is a substantially revised and augmented version of several extended abstracts that introduced the PAA framework~\cite{Marschall2008} and outlined some of the applications presented here~\cite{DBLP:conf/recomb/KaltenbachBR06,Herms2008,Marschall2009,Marschall2010}.

Let us give an overview of the application domains of PAAs considered in this paper.
We begin with the field of pattern matching statistics. Biological sequence analysis is often concerned with the search for structure in long strings like DNA, RNA or amino acid sequences. Frequently, ``search for structure'' means to look for patterns that occur very often. An important point in this process is to sensibly define a notion of ``very often''.
One option is to consult the statistical significance of an event:
Suppose we have found a certain pattern~$\occcount$ times in a given sequence. What is the probability of observing~$\occcount$ or more matches just by chance?
The answer to this question depends on the used null model, \ie the notion of ``by chance''.
It turns out that the PAA framework paves the way to using quite general null models; finite-memory text models as used in this article comprise \iid models, Markovian models of arbitrary order and character-emitting hidden Markov models (HMMs).

The PAA framework can also be applied to the exact analysis of algorithms. Traditionally, best case, average case, and worst case behavior of algorithms are considered. In contrast, we construct PAAs to compute the whole exact distribution of costs of arbitrary window-based pattern matching algorithms like Horspool's or Sunday's algorithm. For these algorithm, we present (perhaps surprising) exemplary results on short patterns and moderate text lengths.

Another application arises when searching for a (biological) query sequence, such as a DNA or protein sequence, in a comprehensive database. The goal is to quickly retrieve all sufficiently similar sequences.
Heuristic methods use so-called \emph{alignment seeds} in order to first detect candidate sequences which are then investigated more carefully. 
To evaluate the quality of such a seed, one computes its sensitivity or hitting probability, \ie the fraction of all desired target sequences it hits. 
Similarly, we ask which fraction of non-related sequences is hit by a seed by chance, as this quantity directly translates into unnecessary subsequent alignment work.
The stochasticity arises from directly modelling alignments of similar (or non-similar) sequences rather than modelling the sequences themselves.
We use the PAA framework to compute the match distribution and, in particular, the sensitivity of certain filtration seeds under finite-memory null models.

Next, we investigate protein identification by mass spectrometric analysis.
If one can describe the typical length and mass of peptide fragments measured by the mass spectrometer, one can define a reliable comparison of so-called peptide mass fingerprints, based on an underlying null model.
In this article, we compute the joint length-mass distribution of random peptide fragments.
Furthermore, we calculate the occurrence probability of fragments in a given mass range, i.e., the probability that cleaving a random protein of a given length yields at least one fragment within this mass range.
This probability aids the interpretation of mass spectra as it gives the significance of a measured peak.
Again, our framework permits to use arbitrary finite-memory protein models.

The final application concerns DNA sequencing: The task of determining a DNA sequence has seen great technological progress over the last decades. 
For a particular sequencing technology (``454 sequencing''), the length of a sequenced DNA fragment depends on the order of its characters.
We compute the exact length distribution of such fragments.
With this distribution we can specify the technology settings that yield the longest reads on average if statistical properties of the genome under consideration are known, improving the sequencing performance by up to 10\%.

Most of the mentioned applications have, in some form or another, previously been discussed in the literature, but never been identified as instances of the same abstract scheme. The contribution of this article is twofold. On the one hand, we introduce a generic framework unifying the view on the presented applications. On the other hand, we show that, in all considered applications, not only known results can be reproduced using PAAs, but new achievements are made. Since the range of applications is quite diverse, we give further references to relevant literature in each section separately.

\paragraph{Organization of the Article.}
In the first part of this article we set up the PAA framework. Specifically, we formally introduce PAAs in Section~\ref{sec:paa} and give generic algorithms in Section~\ref{sec:statevalue}.
We consider waiting time problems on PAAs in Section~\ref{sec:waitingtimes}.
In Section~\ref{sec:paa_on_randseqs}, \emph{deterministic arithmetic automata} and \emph{finite-memory text models} are defined and shown to be a convenient means of specifying a PAA.

Having the framework in place, we explore several application domains. 
Section~\ref{sec:pm_stat} deals with the statistics of patterns on random texts. 
In Section~\ref{sec:horspool}, PAAs are employed for the analysis of window-based pattern matching algorithms.
We cover the field of alignment seed statistics in Section~\ref{sec:seeds}. 
Then, in Section~\ref{sec:fragment_masses}, we apply PAAs to mass statistics of fragments resulting from enzymatic digestion. 
The optimization of read lengths in 454~sequencing is discussed in Section~\ref{sec:sequencing}. 
We conclude the article with a summarizing discussion in Section~\ref{sec:discussion}.

The relations between this article and preliminary versions published in conference proceedings are as follows. Sections~\ref{sec:paa},~\ref{sec:statevalue} and parts of Section~\ref{sec:pm_stat} are based on~\cite{Marschall2008}. Section~\ref{sec:paa_on_randseqs} is based on~\cite{Marschall2010}.  Section~\ref{sec:clump_sizes} contains material from~\cite{Marschall2009}.
Sections~\ref{sec:horspool},~\ref{sec:seeds}, and~\ref{sec:fragment_masses} are based on~\cite{Marschall2010}, \cite{Herms2008}, and~\cite{DBLP:conf/recomb/KaltenbachBR06}, respectively.

An extended version of~\cite{Marschall2010} has been submitted for consideration in a special issue of a journal, a preprint is available as~\cite{Marschall2010b}. There, the analysis of pattern matching algorithms is further generalized and applied to more algorithms. Here, we include a basic example in Section~\ref{sec:horspool} to give another example of the utility of PAAs.

\paragraph{Notation and Conventions.}
The natural numbers (without zero) are denoted~$\N$; to include zero, we write~$\N_0$. Throughout the article, $\Sigma$ is a finite alphabet; as usual, $\Sigma^*$ denotes the set of all finite strings. Indices are zero-based, \ie $s=\chr{s}{0}\ldots\chr{s}{\len{s}-1}$ for $s\in\Sigma^*$.
Substrings, prefixes, and suffixes are written $\substr{s}{i}{j}:=\chr{s}{i}\ldots\chr{s}{j}$, $\prefix{s}{i}:=\chr{s}{0}\ldots\chr{s}{i}$, and $\suffix{s}{i}:=\chr{s}{i}\ldots\chr{s}{\len{s}-1}$, respectively.
All stochastic processes considered in this article are discrete. Therefore, appropriate probability spaces can always be constructed; we do not clutter notation by stating them explicitly. By~$\prob$, we refer to a probability measure; $\dist(X)$ denotes the distribution of the random variable~$X$. Iverson brackets are written $\iverl\cdot\iverr$, \ie $\iverl A\iverr=1$ if the statement~$A$ is true and $\iverl A\iverr=0$ otherwise.

\section{Probabilistic Arithmetic Automata}\label{sec:paa}
In this section, we define \emph{probabilistic arithmetic automata (PAAs)} in order to formalize chains of operations with probabilistic operands. PAAs can be interpreted as generalized Markov Additive Processes (MAP)~\cite{cinlar72markov,cinlar72markova} in the discrete case.

\begin{definition}[Probabilistic Arithmetic Automaton]\label{def:paa} A \emph{probabilistic arithmetic automaton}~$\paa$ is a tuple
\[\paa=\paatuple\,,\]
where
\begin{itemize}
 \item $\stateset$ is a finite set of states,
 \item $\state_0\in\stateset$ is called \emph{start state},
 \item $\paatrans:\stateset\times\stateset\rightarrow [0,1]$ is a transition function with $\sum_{\state'\in \stateset}\paatrans(\state,\state')=1$ for all $\state\in\stateset$, \ie $\big(\paatrans(\state,\state')\big)_{\state,\state'\in\stateset}$ is a stochastic matrix,
 \item $\valset$ is a set called \emph{value set},
 \item $\val_0\in\valset$ is called \emph{start value},
 \item $\emiset$ is a finite set called \emph{emission set},
 \item each $\emidist_\state: \emiset\rightarrow [0,1]$ is an emission distribution associated with state $\state$,
 \item each $\op_\state: \valset\times\emiset\rightarrow\valset$ is an operation associated with state $\state$.
\end{itemize}
\end{definition}

We attach the following semantics: At first, the automaton is in its start state $\state_0$, as for a classical deterministic finite automaton (DFA). 
In a DFA, the transitions are triggered by input symbols. 
In a PAA, the transitions are purely probabilistic; $\paatrans(\state,\state')$ gives the chance of going from state $\state$ to state $\state'$. 
Note that the tuple $(\stateset,\paatrans,\dirac_{\state_0})$ defines a Markov chain on state set $\stateset$ with transition matrix $\paatrans$, where the initial distribution $\dirac_{\state_0}$ is the Dirac distribution assigning probability 1 to $\lbrace \state_0\rbrace$.

While going from state to state, a PAA performs a chain of calculations on a set of values~$\valset$. It starts with value~$\val_0$. Whenever a state transition is made, the entered state, say state~$\state$, generates an emission from~$\emiset$ according to the distribution~$\emidist_\state$. The current value and this emission are then subject to the operation~$\op_\state$, resulting in the next value from the value set~$\valset$. Notice that the Markov chain $(\stateset,\paatrans,\dirac_{\state_0})$, together with the emission set~$\emiset$ and the distributions~$\emidist=(\emidist_\state)_{\state\in\stateset}$, defines a hidden Markov model (HMM). In the context of HMMs, however, the focus usually rests on the sequence of emissions, whereas we are interested in the value resulting from a chain of operations on these emissions.

\begin{figure}
\begin{center}
\includegraphics[width=.35\textwidth]{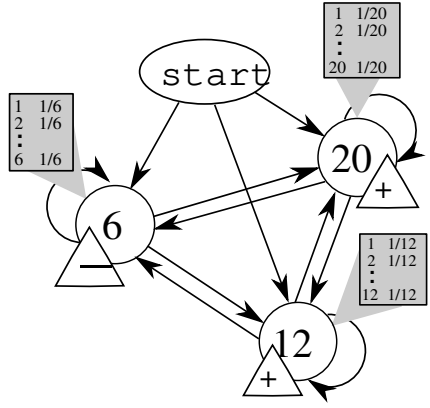}
\end{center}
\caption{Illustration of a PAA for the dice example. Each of the three dice is represented by a state (circles). Emission distributions are depicted as gray boxes. Each arrow stands for a possible state transition, each transition having a probability of~$1/3$. The state's operations are given in triangles next to the state. For the start state, emission distribution and operation are not drawn.}\label{fig:paa_ex}
\end{figure}

By introducing PAAs, we emphasize that many applications can naturally be modelled as a chain of operations whose result is of interest. When compared to Markov chains, PAAs do not offer an increase in expressive power. In fact, from a theoretical point of view, every PAA might be seen as a Markov chain on the state space~$\stateset\times\valset$.
Thus, we advocate PAAs not because of their expressive power but for their merits as a modelling technique. As we shall see, the framework lends itself to many applications and often allows simple and intuitive problem formulations. Before we formalize the introduced semantics in Definition~\ref{def:paa_processes}, let us come back to the dice example from Section~\ref{sec:introduction}.
\begin{example}[Dice]
We model each of the three dice as a PAA state. 
All transition probabilities equal $1/3$ and the emissions have uniform distributions over the number of faces of the respective dice. 
If we are interested in the maximum, each state's operation is the maximum. 
In general, we can associate individual operations with each state, for instance: ``sum the numbers from the 12- and 20-faced dice and subtract the numbers seen on the 6-faced die''.
The value set is $\Z$, the start value is naturally $\val_0=0$.
The corresponding PAA is illustrated in Figure~\ref{fig:paa_ex}.
\end{example}

\begin{definition}[Stochastic processes induced by a PAA]\label{def:paa_processes}
For a given PAA $\paa=\paatuplebrief$, we denote its state process by~$(\stateproc_t^\paa)_{t\in\N_0}$. It is defined to be a Markov chain with $\stateproc_0^\paa\equiv \state_0$ and 
\begin{equation}\label{eqn:state_process}
\begin{split}
   & \prob\left(\stateproc_{t+1}^\paa=\state_{t+1}\given \stateproc_{t}^\paa=\state_t,\ldots, \stateproc_{0}^\paa=\state_0\right) \\
=\ & \prob\left(\stateproc_{t+1}^\paa=\state_{t+1}\given \stateproc_{t}^\paa=\state_t\right) = \paatrans(\state_t,\state_{t+1})
\end{split}
\end{equation}
for all $\state_0,\ldots \state_{t+1}\in \stateset$. 
Further, we define the emission process~$(\emiproc_t^\paa)_{t\in\N_0}$ through
\begin{equation}\label{eqn:emission_process}
\begin{split}
  & \prob\left(\emiproc_t^\paa=\emi\given \stateproc_0^\paa=\state_0,\ldots, \stateproc_t^\paa=\state_t,\emiproc_0^\paa=\emi_0,\ldots, \emiproc_{t-1}^\paa=\emi_{t-1}\right)\\
=\ & \prob\left(\emiproc_t^\paa=\emi\given \stateproc_t^\paa=\state\right)=\emidist_\state(\emi)\,,
\end{split}
\end{equation}
\ie the current emission depends solely on the current state.
We use $(\stateproc_t^\paa)_{t\in\N_0}$ and $(\emiproc_t^\paa)_{t\in\N_0}$ to define the process of values~$(\valproc_t^\paa)_{t\in\N_0}$ resulting from the performed operations:
\begin{equation}\label{eqn:value_process}
\valproc_0^\paa\equiv \val_0\mboxspace{and}\valproc_t^\paa=\op_{\stateproc_t^\paa}\left(\valproc_{t-1}^\paa,\emiproc_t^\paa\right)\mbox{\,.}
\end{equation}
If the considered PAA is clear from the context, we omit the superscript~$\paa$ and write $(\stateproc_t)_{t\in\N_0}$, $(\valproc_t)_{t\in\N_0}$, and $(\emiproc_t)_{t\in\N_0}$, respectively.
\end{definition}

\section{Computing the State-Value Distributions of PAAs}\label{sec:statevalue}
We describe two algorithms to compute the distribution of resulting values. In other words, we seek to calculate the distribution $\dist(\valproc_\totalsteps)$ of the random variable $\valproc_\totalsteps$ for a given $\totalsteps$. 
The idea is to compute the joint distribution $\dist(\stateproc_\totalsteps,\valproc_\totalsteps)$ and then to derive the sought distribution by marginalization:
\begin{equation}
\prob(\valproc_\totalsteps=\val)=\sum_{\state\in \stateset}\prob(\stateproc_\totalsteps=\state,\valproc_\totalsteps=\val)\mbox{\,.}
\end{equation}
For the sake of a shorter notation, we define $f_{t}(\state,\val):=\prob(\stateproc_t=\state,\valproc_t=\val)$ for $t\in\N_0$, $\state\in\stateset$, $v\in\valset$.

A slight complication arises when $\valset$ is infinite.
However, for each $t$, the range of $\valproc_t$ is finite, as it is a function of the states and emissions up to time $t$, and these are finite sets.
We define $\valset_t := \text{range } \valproc_t$ and $\valsize_\totalsteps := \max_{0\leq t\leq \totalsteps}\, |\valset_t|$.
Clearly $\valsize_\totalsteps \leq (|\stateset|\cdot|\emiset|)^\totalsteps$. 
Therefore all actual computations are on finite sets.
As we will see, in many applications, $\valsize_\totalsteps$ grows only polynomially (even linearly) with $\totalsteps$.
In the following, we shall understand $\valset$ as the appropriate union of $\valset_t$ sets. 
Running times of algorithms are given in terms of $\valsize_n$.

\subsection{Basic Algorithm}\label{sec:evo_basic}
We discuss an algorithm to compute the distribution $f_\totalsteps = \dist(\stateproc_\totalsteps,\valproc_\totalsteps)$. 
A basic recurrence relation follows from Definitions~\ref{def:paa} and~\ref{def:paa_processes}.
\begin{lemma}[State-value recurrence]\label{lem:basic_recurrence}
For a given PAA, the state-value distribution can be computed by 
\begin{equation}\label{eqn:basic_recurrence0}
f_0(\state,\val)=
\begin{cases}
1 & \mbox{if }\state=\state_0\mbox{ and }\val=\val_0\,,\\
0 & \mbox{otherwise}\,,
\end{cases}
\end{equation}
and
\begin{equation}\label{eqn:basic_recurrence}
f_{t+1}(\state,\val)=\sum_{\state'\in \stateset}\;\sum_{(\val',\emi)\in\op^{-1}_\state(\val)}f_t(\state',\val')\cdot \paatrans(\state',\state)\cdot \emidist_\state(\emi)\mbox{\,,}
\end{equation}
where $\op^{-1}_\state(\val)$ denotes the inverse image set of $\val$ under $\op_\state$.
\end{lemma}
\begin{proof}
Equation~\eqref{eqn:basic_recurrence0} follows directly from~\eqref{eqn:state_process} and~\eqref{eqn:value_process}. Let us verify Equation~\eqref{eqn:basic_recurrence}:
\begin{align*}
f_{t+1}(\state,\val)&=\prob\left(\stateproc_{t+1}=\state,\valproc_{t+1}=\val\right) \\
&=\sum_{\state'\in \stateset}\sum_{\val'\in\valset}\sum_{\emi\in \emiset}\prob\left(\stateproc_{t+1}=\state,\valproc_{t+1}=\val,\stateproc_t=\state',\valproc_t=\val',\emiproc_{t+1}=\emi\right) \\
&=\sum_{\state'\in \stateset}\sum_{\val'\in\valset}\sum_{\emi\in \emiset}\prob\left(\stateproc_{t+1}=\state,\valproc_{t+1}=\val,\emiproc_{t+1}=\emi\given \stateproc_t=\state',\valproc_t=\val'\right)\cdot f_{t}(\state',\val') \\
&=\sum_{\state'\in \stateset}\sum_{\val'\in\valset}\sum_{\emi\in \emiset}\llbracket\op_\state(\val',\emi)=\val\rrbracket\cdot\prob\left(\stateproc_{t+1}=\state,\emiproc_{t+1}=\emi\given \stateproc_t=\state',\valproc_t=\val'\right)\cdot f_{t}(\state',\val') \\
&=\sum_{\state'\in \stateset}\sum_{(\val',\emi)\in \op_\state^{-1}(\val)}\underbrace{\prob\left(\stateproc_{t+1}=\state,\emiproc_{t+1}=\emi\given \stateproc_t=\state',\valproc_t=\val'\right)}_{(*)}\cdot f_{t}(\state',\val')\,.
\end{align*}
We further evaluate the expression~$(*)$:
\begin{align*}
(*)= &\prob\left(\stateproc_{t+1}=\state,\emiproc_{t+1}=\emi\given \stateproc_t=\state',\valproc_t=\val'\right) \\
= &\underbrace{\prob\left(\emiproc_{t+1}=\emi\given \stateproc_t=\state',\stateproc_{t+1}=\state,\valproc_t=\val'\right)}_{\stackrel{(i)}{=}\emidist_\state(\emi)}\cdot\underbrace{\prob\left(\stateproc_{t+1}=\state\given \stateproc_t=\state',\valproc_t=\val'\right)}_{\stackrel{(ii)}{=}\paatrans(\state',\state)}\,,
\end{align*}
where $(i)$ is true because of~\eqref{eqn:emission_process} and~\eqref{eqn:value_process} and~$(ii)$ follows from the fact that~$(\stateproc_t)_{t\in\N_0}$ is a Markov chain. 
\end{proof}

We start with the distribution $f_0$ and calculate the subsequent distributions by applying Equation~\eqref{eqn:basic_recurrence} until we obtain the desired $f_\totalsteps$. A straightforward implementation of Equation~\eqref{eqn:basic_recurrence} results in a \emph{pull-strategy}; that means each entry in the table representing $f_{t+1}$ is calculated by ``pulling over'' the required probabilities from table $f_t$. Note that this approach makes it necessary to calculate $\op^{-1}$ in a preprocessing step. In order to avoid this, we may implement a \emph{push-strategy}, meaning that we iterate over all entries in $f_t$ rather than $f_{t+1}$ and ``push'' the encountered summands over to the appropriate places in table $f_{t+1}$; in effect, we just change the order of summation. Algorithm~\ref{al:paageneral} shows the push-strategy in detail.

\begin{algorithm}
\caption{\textsc{PaaDist}}
\label{al:paageneral}
\begin{algorithmic}[1]
\vspace*{.1cm}
\REQUIRE $f_{0}=\dist(\stateproc_{0},\valproc_{0})$, $\totalsteps\in \N_0$, 
	space for two tables of size $|\stateset|\times \valsize_\totalsteps$
\ENSURE $f_\totalsteps=\dist(\stateproc_\totalsteps,\valproc_\totalsteps)$
\FOR{$t=1$ to $\totalsteps$}
     \label{line:paageneral_outerloop}
     \STATE initialize $f_t(\state,\val)\equiv 0$ for all $\state\in\stateset$, $\val\in\valset_t$
     \FORALL{$\state\in \stateset$ and $\val\in \valset_{t-1}$} 
          \label{line:paageneral_middleloop}
          \FORALL{$\state'\in \stateset$ and $\emi\in \emiset$} \label{line:paageneral_innerloop}
               \STATE $\val'\gets \op_{\state'}(\val,\emi)$
               \STATE $f_t(\state',\val')\gets f_t(\state',\val')+f_{t-1}(\state,\val)\cdot \paatrans(\state,\state')\cdot \emidist_{\state'}(\emi)$
          \ENDFOR
     \ENDFOR
\ENDFOR
\STATE return $f_\totalsteps$
\end{algorithmic}
\end{algorithm}

In the course of the computation, we have to store two distributions, $f_t$ and $f_{t+1}$, at a time. Once $f_{t+1}$ is calculated, $f_t$ can be discarded. 
Since the table at time~$t$ has a size of $|\stateset|\times|\valset_t|$, the total space consumption is $\OO(|\stateset|\cdot \valsize_\totalsteps)$. 
Computing $f_t$ from $f_{t-1}$ takes $\OO(|\stateset|\cdot |\valset_t| + |\stateset|^2\cdot |\valset_{t-1}| \cdot |\emiset|)$ time, as can be seen from Algorithm~\ref{al:paageneral}. 
We arrive at the following lemma.

\begin{lemma}\label{lem:paa_basic}
Given a PAA $\paatuplebrief$, the distribution of values $\dist(\valproc_\totalsteps)$ can be computed in $\OO(\totalsteps\cdot |\stateset|^2\cdot \valsize_\totalsteps \cdot |\emiset|)$ time and $\OO(|\stateset|\cdot \valsize_\totalsteps)$ space.
\end{lemma}

\subsection{Doubling Technique}\label{sec:evo_doubling}
If $\totalsteps$ is large, executing the above algorithm may be slow.
In this section, we present an alternative algorithm that may be favorable for large $\totalsteps$.
To derive this algorithm, we consider the conditional probability
\begin{equation}
U^{(t)}(\state_1,\state_2,\val_1,\val_2):=\prob\big(\stateproc_{t_0+t}=\state_2,\valproc_{t_0+t}=\val_2\,\big|\,\stateproc_{t_0}=\state_1,\valproc_{t_0}=\val_1\big)\mbox{\,.}
\end{equation}
Note that $U^{(t)}$ does not depend on $t_0$, because transition as well as emission probabilities do not change over ``time'' (a property called \emph{homogeneity}). Once $U^{(\totalsteps)}$ is known, we can simply read off the desired distribution $\dist(\stateproc_\totalsteps,\valproc_\totalsteps)$:
\begin{equation}
\prob(\stateproc_\totalsteps=\state,\valproc_\totalsteps=\val)=U^{(\totalsteps)}(\state_0,\state,\val_0,\val)\mbox{\,.}
\end{equation}
The following lemma shows how $U^{(t)}$ can be computed.
\begin{lemma}\label{lem:paa_doubling}
Let $\paatuplebrief$ be a PAA and $(\stateproc_t)_{t\in\N_0}$ and $(\valproc_t)_{t\in\N_0}$ its state and value process, respectively. Then,
\begin{equation}\label{eqn:doubling_start}
U^{(1)}(\state_1,\state_2,\val_1,\val_2)=\paatrans(\state_1,\state_2)\cdot\sum_{\substack{\emi\in \emiset:\\ \op_{\state_2}(\val_1,\emi)=\val_2}}\emidist_{\state_2}(\emi)
\end{equation}
and, for all $t_1\in\N_0$ and $t_2\in\N_0$,
\begin{equation}\label{eqn:doubling}
U^{(t_1+t_2)}(\state_1,\state_2,\val_1,\val_2)=\sum_{\state'\in \stateset}\sum_{\val'\in\valset}U^{(t_1)}(\state_1,\state',\val_1,\val')\cdot U^{(t_2)}(\state',\state_2,\val',\val_2)\mbox{\,.}
\end{equation}
Using these recurrences, the distribution of values $\dist(\valproc_\totalsteps)$ can be computed in $\OO(\log \totalsteps\cdot|\stateset|^3\cdot \valsize_\totalsteps^3)$ time and $\OO(|\stateset|^2\cdot \valsize_\totalsteps^2)$ space.
\end{lemma}
\begin{proof}
Equation~\eqref{eqn:doubling_start} follows from Definition~\ref{def:paa_processes}, while Equation~\eqref{eqn:doubling} follows from the Chapman-Kolmogorov Equation for homogeneous Markov chains when the PAA is seen as a Markov chain with state space~$\stateset\times\valset$.
Computing $U^{(t_1+t_2)}$ from $U^{(t_1)}$ and $U^{(t_2)}$ takes $\OO(|\stateset|^3\cdot \valsize_\totalsteps^3)$ time, as follows from Equation~\eqref{eqn:doubling}.
On the other hand, one step suffices to obtain $U^{(2t)}$ from $U^{(t)}$. 
Thus, we can compute all $U^{(2^b)}$ for $0\leq b\leq\lceil \log(\totalsteps)\rceil$ in $\lceil \log(\totalsteps)\rceil$ steps, which in turn can be combined into $U^{(\totalsteps)}$ in at most $\lceil \log(\totalsteps)\rceil$ steps.
\end{proof}

We note that the doubling technique is asymptotically faster if $\valsize_\totalsteps$ is $o(\sqrt{n/\log n})$ and $\stateset$ and $\emiset$ are fixed.


\section{Waiting Times}
\label{sec:waitingtimes}
Besides calculating the distribution of values after a fixed number of steps, we can ask for the distribution of the number of steps needed to reach a certain value or a certain state. 
Such \emph{waiting time} problems play an important role in many applications. 
We discuss examples in sections~\ref{sec:pm_stat} and~\ref{sec:sequencing}. 
A classical treatment of waiting time problems is given in \cite{Feller1968}. 
Applications to occurrence problems in texts are reviewed in \cite{ReiSchWat00}.

\begin{definition}[Waiting time for a value]
\label{def:waitingtime}
The waiting time for a set of target values $\mathcal{T}\subset\valset$ is a random variable defined as
$W_\mathcal{T}:=\min\{t\in\N_0 \given \valproc_t\in\mathcal{T}\}$ if this set is not empty,
and defined as infinity otherwise.
\end{definition}

While $\prob\left(W_\mathcal{T}\geq t\right)$ may be nonzero for all $t\in\N$, we are frequently only interested in the distribution up to a fixed time $\tmax$.
Then, of course, the exact values of $\dist(W_\mathcal{T})(t) = \prob(W_\mathcal{T}=t)$ are unknown for $t>\tmax$, but their total probability $\prob(W_\mathcal{T}>\tmax)$ is known, and $\tmax$ is typically chosen such that this total probability remains below a desired threshold.

\begin{lemma}\label{lem:waiting_value}
Let $\paatuplebrief$ be a PAA and $\mathcal{T}\subset\valset$. 
Then, the probabilities $\dist(W_\mathcal{T})(0),$ $\ldots,\dist(W_\mathcal{T})(\tmax)$ can be computed in $\OO\big(\tmax\cdot|\stateset|^2\cdot |\emiset| \cdot (\valsize_{\tmax}-|\mathcal{T}|)\big)$ time and $\OO\big(|\stateset|\cdot (\valsize_{\tmax}-|\mathcal{T}|)\big)$ space.
Alternatively, this can be done using $\OO\big(\log \tmax \cdot |\stateset|^3 \cdot (\valsize_{\tmax} - |\mathcal{T}|)^3 \big)$ time and $\OO\big(|\stateset|^2\cdot(\valsize_{\tmax}-|\mathcal{T}|)^2\big)$ space.
\end{lemma}
\begin{proof}
We construct a modified PAA by defining a new value set $\valset':=(\valset\setminus\mathcal{T})\cup\{\bullet,\circ\}$, assuming (without loss of generality) that $\bullet,\circ\notin\valset$, and new operations
\[ \op'_\state(\val,\emi):= \begin{cases}
  \op_\state(\val,\emi) & \mbox{if }\val\notin\{\bullet,\circ\} \mbox{ and } \op_\state(\val,\emi)\notin\mathcal{T}, \\
  \bullet & \mbox{if } \val\notin\{\bullet,\circ\}\mbox{ and } \op_\state(\val,\emi)\in\mathcal{T}, \\
  \circ & \mbox {if } \val\in\{\bullet,\circ\} \\
\end{cases} 
\]
for all $\state\in\stateset$.
Let $\valproc'_t$ be the modified value process.
Using the modified PAA, the probability of waiting time~$t$ can be expressed as
\[ \prob(W_\mathcal{T}=t)=\prob(\valproc'_t=\bullet). \]
Runtime and space bounds follow from Lemmas~\ref{lem:paa_basic} and~\ref{lem:paa_doubling}.
\end{proof}

Besides waiting for a set of values, we may also wait for a set of states.
Since a PAA's state process does not depend on emission and value processes, the remainder of this section solely concerns the Markov chain $(\stateset,\paatrans,\dirac_{\state_0})$, which is part of the PAA $\paatuplebrief$.
Waiting times in Markov chains are a well-studied topic with particular interest in pattern occurrences~\cite{ReiSchWat00} and queuing
theory~\cite{bremaud1999}.
For completeness and implementation purposes within the PAA framework, we briefly restate the construction here.

\begin{definition}[Waiting time for a state]
The waiting time for a set of target states $\mathcal{S}\subset\stateset$ is a random variable defined as
$W_\mathcal{S}:=\min\{t\in\N_0 \given \stateproc_t\in\mathcal{S}\}$ if this set is not empty
and defined as infinity otherwise.
\end{definition}

\begin{lemma}\label{lem:waiting_state}
Let $\paatuplebrief$ be a PAA, $\alpha:\stateset\to[0,1]$ be a probability distribution on~$\stateset$, and $\mathcal{S}\subset\stateset$ be a set of target states.
Consider the Markov chain $(\stateset,\paatrans,\alpha)$, let $(\stateproc'_t)_{t\in\N_0}$ be its state process,
and let $W'_\mathcal{S}:=\min\{t\in\N_0 \given \stateproc'_t\in\mathcal{S}\}$ be the waiting time for states $\mathcal{S}$.
Then $\dist(W'_\mathcal{S})(0),\ldots,\dist(W'_\mathcal{S})(\tmax)$ can be computed in $\OO(\tmax \cdot |\stateset|^2)$ time and $\OO(|\stateset|)$ space, or in $\OO(\log \tmax \cdot |\stateset|^3)$ time and $\OO(|\stateset|^2)$ space using the doubling technique.
If $\alpha=\dirac_{\state_0}$, then $W_\mathcal{S}=W'_\mathcal{S}$.
\end{lemma}
\begin{proof}
As in the proof of Lemma~\ref{lem:waiting_value}, we introduce an aggregation state $\bullet$ to replace $\mathcal{S}$ and an absorbing state $\circ$ to ``flush'' $\bullet$. Then $\prob(W'_\mathcal{S}=t)=\prob(\stateproc'_t=\bullet)$.
\ignore{
Consider the modified state space~$\stateset':=(\stateset\setminus\mathcal{S})\cup\{\bullet,\circ\}$. Define a transition function $T':\stateset'\to\stateset'$ through
\[T'(\state,\state'):=
\begin{cases}
T(\state,\state') & \mbox{if }\state,\state'\in\stateset\setminus\mathcal{S}\,, \\
\sum_{\state''\in\mathcal{S}}T(\state,\state'') & \mbox{if }\state\in\stateset\setminus\mathcal{S}\mbox{ and }\state'=\bullet\,, \\ 
1 & \mbox{if }\state\in\{\bullet,\circ\}\mbox{ and }\state'=\circ\,, \\
0 & \mbox{otherwise}\,.
\end{cases}\]
Now, the waiting time can be written
\[\prob(W_\mathcal{S}=t)=\prob(\stateproc'_t=\bullet)\,.\]
The claimed time and space bounds follow from the fact that $\dist(\stateproc_{t+1})$ can be computed from $\dist(\stateproc_{t})$ in $\OO(|\stateset|^2)$ time using the recurrence
\[\prob(\stateproc_{t+1}=\state')=\sum_{\state\in\stateset}\prob(\stateproc_{t}=\state)\cdot\paatrans(\state,\state')\,.\]
If $\alpha=\dirac_{\state_0}$, then $\stateproc_t=\stateproc'_t$ for all $t\in\N_0$ by Definition~\ref{def:paa_processes} and therefore $W_\mathcal{S}=W'_\mathcal{S}$.
}
\end{proof}

When $\alpha=\dirac_{\state_0}$, the above lemma yields the waiting time for the first event of reaching one of the states in $\mathcal{S}$. 
We further consider the waiting time of a return event 
\[W^{t_0}_\mathcal{S}:=\min\{t\in\N \given \stateproc_{t_0+t}\in\mathcal{S}\}.\]
If the Markov chain is aperiodic and irreducible, it has a unique stationary state distribution, against which the PAA state distribution converges exponentially fast. 
We can then use Lemma~\ref{lem:waiting_state} to compute
\[\lim_{t\to\infty}\prob\big(W^{t}_\mathcal{S}=t'\ \big|\ \stateproc_{t}\in\mathcal{S}\big)
\quad\text{for each } t'\in\N
\]
by choosing $\alpha$ in Lemma~\ref{lem:waiting_state} as the stationary distribution restricted to~$\mathcal{S}$.
 

\section{PAAs Based on Random Sequences}\label{sec:paa_on_randseqs}
We discuss the construction of PAAs modelling the deterministic processing of random sequences. That means we assume to be given a mechanism that processes sequences character by character and deterministically computes a value for a given string. A pattern matching algorithm that computes the number of matches in a given sequence might serve as an example. In this section, we ask for the distribution of resulting values when a \emph{deterministic} computation is applied to \emph{random} strings. Therefore we first define \emph{text models} and \emph{deterministic arithmetic automata} to represent random texts and deterministic computations, respectively, and then combine both into a PAA.

\subsection{Random Text Models}\label{sec:text_model}
Given an alphabet~$\Sigma$, a random text is a stochastic process~$(S_t)_{t\in\N_0}$, where each $S_t$ takes values in~$\Sigma$. A text model $\prob$ is a probability measure assigning probabilities to (sets of) strings. 
It is given by (consistently) specifying the probabilities $\prob(S_0\ldots S_{\len{s}-1}=s)$ for all~$s\in\Sigma^\ast$. 
We only consider finite-memory models in this article which are formalized in the following definition.

\begin{definition}[Finite-memory text model]\label{def:text_model}
A \emph{finite-memory text model} is a tuple $(\mathcal{C},c_0,\Sigma,\varphi)$, where $\mathcal{C}$ is a finite state space (called \emph{context space}), $c_0\in\mathcal{C}$ a start context, $\Sigma$ an alphabet, and $\varphi: \mathcal{C}\times\Sigma\times\mathcal{C}\to[0,1]$ with $\sum_{\sigma\in\Sigma,c'\in\mathcal{C}}\varphi(c,\sigma,c')=1$ for all $c\in\mathcal{C}$. The random variable giving the text model state after $t$~steps is denoted~$C_t$ with $C_0:\equiv c_0$.
A probability measure is now induced by stipulating
\[ \prob(S_0\ldots S_{\totalsteps-1}=s,C_1=c_1,\ldots,C_\totalsteps=c_\totalsteps) 
  := \prod_{i=0}^{\totalsteps-1}\, \varphi(c_{i},\chr{s}{i},c_{i+1})
\]
for all $\totalsteps\in\N_0$, $s\in\Sigma^n$, and $(c_1,\ldots,c_\totalsteps)\in\mathcal{C}^\totalsteps$.
\end{definition}

The idea is that the model given by $(\mathcal{C},c_0,\Sigma,\varphi)$ generates a random text by moving from context to context and emitting a character at each transition, where $\varphi(c,\sigma,c')$ is the probability of moving from context~$c$ to context~$c'$ and thereby generating the letter~$\sigma$.

Note that the probability $\prob(S_0\ldots S_{\len{s}-1}=s)$ is obtained by marginalization over all context sequences that generate $s$.
This can be efficiently done, using the decomposition of the following lemma.
\begin{lemma}\label{lem:text_model}
Let $(\mathcal{C},c_0,\Sigma,\varphi)$ be a finite-memory text model. Then, 
\[ \prob(S_0\ldots S_{\totalsteps}=s\sigma,C_{\totalsteps+1}=c)
   = \sum_{c'\in\mathcal{C}}\, \prob(S_0\ldots S_{\totalsteps-1}=s,C_\totalsteps=c') \cdot \varphi(c',\sigma,c)
\]
for all $n\in\N_0$, $s\in\Sigma^n$, $\sigma\in\Sigma$ and $c\in\mathcal{C}$.
\end{lemma}
\begin{proof}
We have
\begin{align*}
 & \prob(S_0\ldots S_{\totalsteps}=s\sigma,C_{\totalsteps+1}=c) \\
=& \sum_{c_1,\ldots,c_\totalsteps}\, \prob(S_0\ldots S_{\totalsteps}=s\sigma, C_1=c_1,\ldots,C_{\totalsteps}=c_\totalsteps,C_{\totalsteps+1}=c) \\
=& \sum_{c_1,\ldots,c_\totalsteps}\, \prod_{i=0}^{\totalsteps-1}\, \varphi(c_{i},\chr{s}{i},c_{i+1})\cdot\varphi(c_\totalsteps,\sigma,c) \\
=& \sum_{c_{\totalsteps}\in\mathcal{C}}\, 
   \left(\sum_{c_1,\ldots,c_{\totalsteps-1}}\, \prod_{i=0}^{\totalsteps-1}\,   
   \varphi(c_{i},\chr{s}{i},c_{i+1})\right) \cdot \varphi(c_\totalsteps,\sigma,c) \\
=& \sum_{c_{\totalsteps}\in\mathcal{C}}\prob(S_0\ldots S_{\totalsteps-1}=s,C_{\totalsteps}=c_n)  \cdot\varphi(c_\totalsteps,\sigma,c)\,.
\end{align*}
Renaming $c_n$ to $c'$ yields the claimed result.
\end{proof}

Similar text models are used in~\cite{Kucherov2006}, where they a called probability transducers. 
In the following, we refer to a finite-memory text model $(\mathcal{C},c_0,\Sigma,\varphi)$ simply as text model, as all text models considered in this article are special cases of Definition~\ref{def:text_model}.

For an \iid model, we set $\mathcal{C}=\{\emptystring\}$ and $\varphi(\emptystring,\sigma,\emptystring)=p_\sigma$ for each~$\sigma\in\Sigma$, where~$p_\sigma$ is the occurrence probability of letter~$\sigma$ (and $\emptystring$ may be interpreted as an empty context).
For a Markovian text model of order~$r$, the distribution of the next character depends only on the~$r$ preceding characters (fewer at the beginning); thus we set $\mathcal{C}:=\bigcup_{i=0}^{r}\Sigma^i$. 
The conditional follow-up probabilities are given by
\begin{align*}
\prob(S_i=\chr{s}{i}& \,\given\, S_{i-1}=\chr{s}{i-1},\ldots,S_0=\chr{s}{0}) \\
                    &=
\begin{cases}
\varphi(\prefix{s}{i-1},\chr{s}{i},\prefix{s}{i}) & \mbox{if }i< r\,, \\
\varphi(\substr{s}{i-r}{i-1},\chr{s}{i},\substr{s}{i-r+1}{i}) & \mbox{if }i\geq r\,. \\
\end{cases}
\end{align*}
This notion of text models also covers variable order Markov chains as introduced in~\cite{Schulz2008}, which can be converted into equivalent models of fixed order. Text models as defined above have the same expressive power as character-emitting HMMs, that means, they allow to construct the same probability distributions. For a given HMM, we can construct an equivalent text model by using the same state space (contexts) and setting $\varphi(c,\sigma,c'):=T(c,c')\cdot \emidist_{c'}(\sigma)$, where~$T$ and~$\emidist_{c'}$ are the HMM's transition function and emission distribution attached to state~$c'$, respectively. When, on the other hand, a text model~$(\mathcal{C},c_0,\Sigma,\varphi)$ is given, we construct an equivalent HMM by using $\mathcal{C}^2$ as state space and setting \[T\big((c_1,c_2),(c'_1,c'_2)\big):=
\begin{cases}
\sum_{\sigma\in\Sigma}\varphi(c_2,\sigma,c'_2) & \mbox{if }c_2=c'_1, \\
0 & \mbox{otherwise},
\end{cases}
\]
and 
\[\emidist_{(c_1,c_2)}(\sigma):=\varphi(c_1,\sigma,c_2).\]

\subsection{Deterministic Arithmetic Automata (DAAs)}
In order to model deterministic calculations on sequences, we define a deterministic counter-part to PAAs.
\begin{definition}[Deterministic Arithmetic Automaton, DAA]\label{def:daa}
A \emph{deterministic arithmetic automaton} is a tuple
\[\daa = \big( \stateset, \state_0, \Sigma, \delta, \valset, \val_0, \emiset, (\daaemi_\state)_{\state\in \stateset},(\op_\state)_{\state\in \stateset} \big),\] 
where~$\stateset$ is a finite set of states, $\state_0\in \stateset$ is the start state, 
$\Sigma$ is a finite alphabet, $\delta: \stateset\times\Sigma\to \stateset$ is called transition function, 
$\valset$ is a set of values, $\val_0\in\valset$ is called the start value,
$\emiset$ is a finite set of emissions, $\daaemi_\state\in \emiset$ is the emission associated to state~$\state$, 
and $\op_\state: \valset\times \emiset\to\valset$ is a binary operation associated to state~$\state$.
Further, we define the associated \emph{joint transition function}
\[ \deltajoint: (\stateset\times\valset)\times \Sigma \to (\stateset\times\valset), \qquad
          \deltajoint \big((\state,\val),\sigma\big) := \big(\delta(\state,\sigma)\,,\, \op_{\delta(\state,\sigma)}(\val,\daaemi_{\delta(\state,\sigma)})\big).
\]
We extend the definitions of $\delta$ and $\deltajoint$ inductively to~$\Sigma^*$ in their second argument by setting
\begin{align*}
\delta (\state, \emptystring) &:= \state &&\text{ for the empty string }\emptystring,\\
\delta (\state, x\sigma)      &:= \delta (\delta(\state,x), \sigma) &&\text{ for all~$x\in\Sigma^*$ and~$\sigma\in\Sigma$}, \\
\deltajoint \big((\state,\val),\emptystring\big) &:= (\state,\val),\\
\deltajoint \big((\state,\val),x\sigma\big) &:= \deltajoint \big(\deltajoint((\state,\val),x),\sigma\big).
\end{align*}

When~$\deltajoint \big((\state_0,\val_0),s\big) = (\state,\val)$ for some $\state\in \stateset$ and $s\in\Sigma^*$, we say that~$\daa$ computes value~$\val$ for input~$s$ and define~$\daaval_{\daa}(s) := \val$.
\end{definition}

Informally, a DAA starts with the state-value pair~$(\state_0,\val_0)$ and reads a sequence of symbols from~$\Sigma$.
Being in state~$\state$ with value~$\val$, upon reading~$\sigma\in\Sigma$, the DAA performs a state transition to~$\state':=\delta(\state,\sigma)$ and updates the value to~$\val':=\op_{\state'}(\val,\daaemi_{\state'})$ using the operation and emission of the new state~$\state'$.

For each state~$\state$, the emission~$\daaemi_\state$ is fixed and could be dropped from the definition of DAAs. 
In fact, one could also dispense with values and operations entirely and define a DFA over state space
$\stateset\times\valset$, performing the same operations as a DAA. 
However, we intentionally include values, operations, and emissions to emphasize the connection to PAAs.

\subsection{Constructing PAAs from DAAs and Text Models}
We now formally state how to convert a DAA into a (restricted) PAA, where each emission distribution
is deterministic (assigning probability~$1$ to a particular value), given a text model.

\begin{lemma}[DAA $+$ Text model $\to$ PAA]
\label{lem:daapaa}
Let $(\mathcal{C},c_0,\Sigma,\varphi)$ be a text model and $\daa = \big(\stateset^\daa, \state_0^\daa, \Sigma, \delta, \valset, \val_0, \emiset, (\daaemi_\state)_{\state\in \stateset^\daa}, (\op_\state^\daa)_{\state\in \stateset^\daa} \big)$ be a DAA. 
Then, define
\begin{itemize}
 \item a state space $\stateset:=\stateset^\daa\times\mathcal{C}$,
 \item a start state~$\state_0:=(\state_0^\daa,c_0)$,
 \item transition probabilities
 \begin{equation}\label{eqn:daa_paa_transfunc}
 \paatrans \big( (\state,c),(\state',c')\big) 
   :=  \sum_{\sigma\in\Sigma:\,\delta(\state,\sigma)=\state'}\, 
       \varphi(c,\sigma,c'),
 \end{equation}
 \item (deterministic) emission probability vectors
   \[ \emidist_{(\state,c)}(\emi):=
       \begin{cases}
       1 & \mbox{if } \emi=\daaemi_\state\,, \\
       0 & \mbox{otherwise}\,,
       \end{cases}
   \]
   for all $(\state,c)\in \stateset$.
 \item operations $\op_{(\state,c)}(\val,\emi) := \op^\daa_\state(\val,\emi)$ for all $(\state,c)\in \stateset$.
\end{itemize}
Then, $\paa = \paatuple$ is a PAA with
\[ \dist(\valproc_t) = \dist\big(\daaval_{\daa}(S_{0}\dots S_{t-1})\big)
\]
for all $t\in\N_0$, where~$S$ is a random text according to the text model $(\mathcal{C},c_0,\Sigma,\varphi)$. 
\end{lemma}
\begin{proof}
$\paa$ is a PAA by Definition~\ref{def:paa}.
As in Section~\ref{sec:statevalue}, we define $f_{t}(\state,\val):=\prob(\stateproc_t=\state,\valproc_t=\val)$.
To prove $\dist(\valproc_{t}) = \dist\big(\daaval_{\daa}(S_{0}\dots S_{{t}-1})\big)$, we show that 
\begin{equation}\label{eqn:paadaa}
f_{t}\big((\state^\daa,c),\val\big) = \sum_{s\in\Sigma^{t}}\big\iverl \deltajoint\big((\state_0^\daa,\val_0),s\big)=(\state^\daa,\val)\big\iverr\cdot\prob(S_0\ldots S_{t-1}=s,C_t=c)
\end{equation}
for all $\state^\daa\in \stateset^\daa$, $c\in\mathcal{C}$, $\val\in\valset$, and $t\in\N_0$. For~$t=0$, Equation~\eqref{eqn:paadaa} is correct by definitions of PAAs, DAAs and text models. For $t>0$ we prove it inductively. Assume~\eqref{eqn:paadaa} to be correct for all $t'$ with $0\leq t'<t$.
\allowdisplaybreaks
\begin{align}
  & f_{t}\big(\underbrace{(\state^\daa,c)}_{=:\state},\val\big)  \label{eqn:paadaa1} \\
= & \sum_{\state'\in \stateset}\; \sum_{(\val',\emi)\in\op^{-1}_\state(\val)}\, 
     f_{t-1}(\state',\val')\cdot \paatrans(\state',\state)\cdot \emidist_\state(\emi)  \label{eqn:paadaa2}\\
= & \sum_{\state'\in \stateset}\; \sum_{(\val',\emi)\in\valset\times \emiset}\, 
    \big\iverl\op^\daa_{\state^\daa}(\val',\emi)=\val\big\iverr \cdot f_{t-1}(\state',\val')
    \cdot \paatrans(\state',\state)\cdot\big\iverl\daaemi_{\state^\daa}=\emi\big\iverr \label{eqn:paadaa3}\\
\begin{split}
= & \sum_{\state'^\daa\in \stateset^\daa}\; \sum_{c'\in\mathcal{C}}\; \sum_{(\val',\emi)\in\valset\times \emiset}\,
    \big\iverl\op^\daa_{\state^\daa}(\val',\emi)=\val\big\iverr  \cdot \big\iverl\daaemi_{\state^\daa}=\emi\big\iverr 
    \cdot f_{t-1}(\state',\val')\\
    & \qquad \cdot \sum_{\sigma\in\Sigma}\, \big\iverl\delta(\state'^\daa,\sigma)=\state^\daa\big\iverr
    \cdot \varphi(c',\sigma,c)
\end{split}\label{eqn:paadaa4}\\
\begin{split}
= & \sum_{s\in\Sigma^{t-1}}\; \sum_{\sigma\in\Sigma}\; \sum_{\state'^\daa\in \stateset^\daa}\;
    \sum_{c'\in \mathcal{C}}\; \sum_{(\val',\emi)\in\valset\times \emiset}\,
    \big\iverl \op^\daa_{\state^\daa}(\val',\emi)=\val\big\iverr \cdot 
    \big\iverl \daaemi_{\state^\daa}=\emi\big\iverr \\
& \qquad \cdot \big\iverl \delta(\state'^\daa,\sigma)=\state^\daa\big\iverr \cdot
    \big\iverl \deltajoint\big((\state_0^\daa,\val_0),s\big)=(\state'^\daa,\val')\big\iverr \\
& \qquad \cdot \prob(S_0\ldots S_{t-2}=s,C^{t-1}=c') \cdot \varphi(c',\sigma,c)
\end{split}  \label{eqn:paadaa5}\\
\begin{split}
= & \sum_{s\sigma\in\Sigma^t}\; \sum_{\state'^\daa\in \stateset^\daa}\;
    \sum_{(\val',\emi)\in\valset\times \emiset}\, 
    \big\iverl \op^\daa_{\state^\daa}(\val',\emi)=\val\big\iverr  \cdot
    \big\iverl \daaemi_{\state^\daa}=\emi\big\iverr \cdot
    \big\iverl \deltajoint\big((\state_0^\daa,\val_0),s\big)=(\state'^\daa,\val')\big\iverr \\
& \qquad \cdot \big\iverl \delta(\state'^\daa,\sigma)=\state^\daa\big\iverr
    \cdot \prob(S_0\ldots S_{t-1}=s\sigma,C_t=c)
\end{split}\label{eqn:paadaa6}\\
= & \sum_{s\sigma\in\Sigma^t}\, 
    \big\iverl \deltajoint\big((\state_0^\daa,\val_0),s\sigma\big)=(\state^\daa,\val)\big\iverr \cdot
    \prob(S_0\ldots S_{t-1}=s\sigma,C_t=c)  \label{eqn:paadaa7}
\end{align}
In the above derivation, step \eqref{eqn:paadaa1}$\to$\eqref{eqn:paadaa2} follows from~\eqref{eqn:basic_recurrence}.
Step \eqref{eqn:paadaa2}$\to$\eqref{eqn:paadaa3} follows from the definitions of $\op_\state$ and $\emidist_\state$.
Step \eqref{eqn:paadaa3}$\to$\eqref{eqn:paadaa4} uses the definitions of~$\paatrans$ and $\stateset$ in Lemma~\ref{lem:daapaa}.
Step \eqref{eqn:paadaa4}$\to$\eqref{eqn:paadaa5} uses the induction assumption.
Step \eqref{eqn:paadaa5}$\to$\eqref{eqn:paadaa6} uses Lemma~\ref{lem:text_model}.
The final step \eqref{eqn:paadaa6}$\to$\eqref{eqn:paadaa7} follows by combining the four Iverson brackets summed over $\state'^\daa$ and $(\val',\emi)$ into a single Iverson bracket.
\end{proof}
\begin{remark}
In the above lemma, states having zero probability of being reached from $\state_0$ may be omitted from $\stateset$ and $\paatrans$.
\end{remark}
\begin{lemma}[PAA from DAA; Construction time and space]\ \\
\label{lem:daapaa:time}
\vspace{-1em}
\begin{enumerate}
\item
For a PAA constructed according to Lemma~\ref{lem:daapaa}, the value distribution $\dist(\valproc_\totalsteps)$, or the joint state-value distribution, can be computed with $\OO(\totalsteps \cdot |\stateset^\daa| \cdot |\Sigma| \cdot |\mathcal{C}|^2 \cdot \valsize_\totalsteps)$ operations using $\OO(|\stateset^\daa| \cdot |\mathcal{C}| \cdot \valsize_\totalsteps)$ space.
The same statement holds for computing the waiting time distribution up to time~$\tmax$.
\item
If for all $c\in\mathcal{C}$ and $\sigma\in\Sigma$, there exists at most one $c'\in\mathcal{C}$ such that $\varphi(c,\sigma,c')>0$, then the time is bounded by $\OO(\totalsteps \cdot |\stateset^\daa| \cdot |\Sigma| \cdot |\mathcal{C}| \cdot \valsize_\totalsteps)$.
\item
Using the doubling technique, the distributions can be computed in $\OO(\log \totalsteps \cdot |\stateset^\daa|^3 \cdot |\mathcal{C}|^3 \cdot \valsize_\totalsteps^3)$ time and $\OO(|\stateset^\daa|^2 \cdot |\mathcal{C}|^2 \cdot \valsize_\totalsteps^2)$ space.
\end{enumerate}
\end{lemma}
\begin{proof}\ \\
\vspace{-1em}
\begin{enumerate}
\item
From Lemma~\ref{lem:paa_basic}, we obtain bounds for time and space complexity of $\OO(\totalsteps \cdot |\stateset|^2 \cdot \valsize_\totalsteps \cdot |\emiset|)$ and $\OO(|\stateset|\cdot \valsize_\totalsteps)$, respectively. By construction, $|\stateset| \leq |\stateset^\daa| \cdot |\mathcal{C}|$. Recall that Lemma~\ref{lem:paa_basic} is based on Algorithm~\ref{al:paageneral}. The loops in lines~\ref{line:paageneral_outerloop} and~\ref{line:paageneral_middleloop} together account for a factor of $\OO(\totalsteps \cdot|\stateset|\cdot \valsize_\totalsteps)$ in the time complexity. A factor of $\OO(|\stateset|\cdot|\emiset|)$ is caused by the inner loop in line~\ref{line:paageneral_innerloop}. However, since the constructed PAA has deterministic (\ie Dirac distributed) emissions, we do not need to iterate over all $\emi\in\emiset$ and save a factor of~$|\emiset|$. Furthermore, we only need to iterate over all states reachable in one step. For each $\state\in\stateset$, there exist at most $|\Sigma| \cdot |\mathcal{C}|$ such states by construction of the PAA. Therefore, the inner loop in line~\ref{line:paageneral_innerloop} can be modified to take $\OO(|\Sigma| \cdot |\mathcal{C}|)$ time, yielding the claimed runtime bound.
\item
If for all $c\in\mathcal{C}$ and $\sigma\in\Sigma$, there exists at most one $c'\in\mathcal{C}$ such that $\varphi(c,\sigma,c')>0$, then at most $|\Sigma|$ different states are reachable from each state $\state\in\stateset$. The claimed runtime follows by the same arguments as above.
\item
Alternative time and space complexities for the doubling algorithm follow directly from Lemma~\ref{lem:paa_doubling}.
\end{enumerate}%
\end{proof}

This concludes the derivation of the PAA framework and construction methods.
The following sections are devoted to several applications.



\section{Pattern Matching Statistics}\label{sec:pm_stat}
One application of the introduced framework are pattern matching statistics. An algorithm that searches for a pattern~$\pattern$ maps strings to the number of matches. That means, it deterministically processes a string and, by doing so, computes a value. In this section, we ask for the distribution of the number of occurrences of a given pattern in a random text.

In computational biology, one searches for patterns that occur often and hypothesizes that these patterns carry biological meaning. 
Mere abundance, however, does not necessarily imply that the found pattern is meaningful. 
Short patterns like \texttt{AC} and degenerate patterns like \texttt{ANNNNNNC} will naturally occur quite often in a given stretch of DNA (here \texttt{N} is a wildcard character meaning aNy nucleotide). 
A better approach to quantify a pattern's overrepresentation is to consult the statistical significance: 
Suppose we have found a certain pattern $\occcount$~times in a given sequence. 
What is the probability of observing $\occcount$ or more matches just by chance? 
Precisely this question can be answered by computing the distribution of occurrence counts. 
Given a suitable null model, a procedure to compute the significance of a pattern is a powerful tool in the context of motif discovery, as it allows the comparison of different patterns regardless of their structure and length. 

There are many different types of patterns that are relevant in computational biology, such as single strings, sets of strings, Prosite patterns\footnote{used in the Prosite database (see Hulo et~al.\ \cite{HulBaiBui06}).
A syntax description can be found under\\\url{http://www.expasy.org/tools/scanprosite/scanprosite-doc.html}.}, consensus strings together with a distance measure and a distance threshold, abelian patterns, position weight matrices in connection with a threshold, etc. All these pattern types may be seen as ways to concisely describe a finite sets of strings. Thus, all these patterns may be expressed in the form of deterministic finite automata (DFAs) that recognize the respective string set. As our method is based on this DFA representation, it is very general and flexible regarding the pattern type.

Besides specifying a pattern, one has to decide how overlaps are to be handled. We refer to the used strategy as \emph{counting scheme}. The easiest case is to disallow overlaps at all; we call this scheme \emph{non-overlapping count}. Consequently, we define the \emph{overlapping count} to be the number of substrings that match the given pattern. In the case of a set of strings without any restrictions, this scheme makes counting more complicated as some words may be substrings of others. Many authors avoid the problem---at least partly---by simply counting the positions where at least one pattern ends, which we refer to as \emph{match position count}.

\subsection{Related Work and New Results}
The topic of statistics of words on random texts has been studied extensively. An overview is provided in the book by Lothaire \cite{Lot05}. Chapter~6 (``Statistics on Words with Applications to Biological Sequences''), which is particularly interesting in this context, is based on the overview article by Reinert~et~al.\ \cite{ReiSchWat00}.

In many approaches, a generating function is derived for the sought quantity. Then, typically using symbolic Taylor expansion, the concrete values can be computed. This procedure is, for instance, described by R\'{e}gnier~\cite{Reg00}, who gives formulas for mean, variance and higher statistical moments of the exact occurrence count distribution. 
This approach has the advantage of additionally allowing asymptotic analysis. Her framework is general enough to admit Markovian sources as well as finite sets of patterns to be treated in the overlapping as well as in the non-overlapping case. Closely related is the approach of Nicod\`{e}me~et~al.\ \cite{NicSalFla02}, who present an algorithmic chain to compute the distribution of the match position count for regular expressions.
Lladser~et~al.\ \cite{LlaBetKni08} recently reviewed the field. Their main concern is to bring together involved concepts in a consistent and rigorous manner. They make the connection to the classical field of automata theory and pattern matching explicit and therefore speak of \emph{probabilistic pattern matching}. Furthermore, they describe the relation between finite automata and Markov chains in terms of the \emph{Markov chain embedding} technique. Related approaches to compute the exact p-values based on automata are developed in~\cite{Boeva2007,Nuel2008}. Another dynamic programming approach was  presented in~\cite{ZhaJiaLi07}. It is used to compute exact p-values for position weight matrices describing transcription factor binding sites (TFBS).

In this section, we provide a unifying framework for the efficient computation of pattern matching statistics. In contrast to existing approaches, overlaps are handled correctly and arbitrary finite-memory text models (including HMMs) can be used. 

\subsection{Constructing DAAs from DFAs}\label{subsec:dfa2daa}
As usual, we define a deterministic finite automaton (DFA) to be a tuple $(\stateset,\Sigma,\delta,\state_0,\mathcal{F})$, where $\stateset$ is a finite state space, $\Sigma$ is a finite alphabet, $\delta:\stateset\times\Sigma\to \stateset$ is a transition function, $\state_0$ is a start state, and $\mathcal{F}\subset\stateset$ is a set of accepting states.
Again, we extend $\delta$ to $\Sigma^*$ in its second argument, as for DAAs in Definition~\ref{def:daa}.
Suppose a pattern is given in the form of a DFA, that means, a string is an instance of the pattern if it is accepted by the DFA. We seek to calculate the distribution of the number of occurrences of this pattern. 

To use a DFA for pattern matching, that is, to find all instances of a pattern in a (long) text, we construct an automaton that accepts not only all strings matching the pattern, but all strings that have a suffix matching the pattern (see~\cite{NavRaf02}).
For a pattern given in the form of a DFA or non-deterministic finite automaton (NFA), it is always possible to construct a DFA for this task as follows.
First, we add a self-transition labeled with the whole alphabet $\Sigma$ to the automaton's start state to ensure that the start state remains active all the time.
The result is an NFA, which can be made deterministic again by employing the classical subset construction (explained for example in~\cite{NavRaf02}).
As we see in Sections~\ref{sec:aho_corasick} to~\ref{sec:prosite}, there are often simpler and more direct ways to build the sought DFA.

When a DFA reads a text, it is in an accepting state whenever (at least) one instance of the pattern ends. The number of these events equals the \emph{match position count} as defined above. 
To count the number of times a DFA $(\stateset,\Sigma,\delta,\state_0,\mathcal{F})$ is in an accepting state, we define 
emissions
\[\daaemi_\state:=
\begin{cases}
1 & \mbox{if }\state\in \mathcal{F},\\
0 & \mbox{otherwise}.
\end{cases}
\]
We call a tuple $(\stateset,\Sigma,\delta,\state_0,(\daaemi_\state)_{\state\in\stateset})$ \emph{counting DFA}. 

To compute the distribution of the \emph{overlapping count} instead of the match position count, we have to take into account that more than one match can end at a position in the text. 
For each state, we therefore need to know how many matches end when it is entered. 
As we will see, this is never a problem when the pattern represents a finite set of strings%
\footnote{If the pattern represents an infinite set, such as a regular expression containing a star operator, there may be states in~$\mathcal{F}$ that can be entered when $i$ or $j$ matches end for $i\neq j$, or for which the number of overlapping matches to be counted might be unbounded. In this case, one is essentially restricted to the match position count. Since we only consider finite sets, this is not an issue here.}.
Formally, we define emissions $\daaemi=(\daaemi_\state)_{\state\in\stateset}$, where $\daaemi_\state\in\N_0$ gives the number of matches to be counted upon entering state~$\state$. 

To obtain the \emph{nonoverlapping count}, the automaton can be modified accordingly:
We change the outgoing transitions of each accepting state of the match position count automaton to act as if they originate from the start state $\state_0$.
Therefore, we define a modified transition function $\delta'$ by $\delta'(\state,\sigma) := \delta(\state,\sigma)$ if $q\notin F$, and $\delta'(\state,\sigma) := \delta(\state_0,\sigma)$ otherwise.

As described above, all three counting schemes can be realized by counting DFAs when the pattern is a finite set of strings.
The main idea now is to construct a DAA from this counting DFA by adding the emissions generated in each state, and turn this into a PAA using Lemma~\ref{lem:daapaa}. 
For practical computations, it is often sufficient to truncate the values at a constant $M\in\N$.
(Note that the match position count is always bounded by the length of the processed text, so $\valsize_n = \Theta(n)$.) 
The proof of the following theorem contains the details of the DFA$\to$DAA$\to$PAA construction.

\begin{theorem}
Let a counting DFA $D=(\stateset,\Sigma,\delta,\state_0,(\daaemi_\state)_{\state\in\stateset})$ and a text model $(\mathcal{C},c_0,\Sigma,\varphi)$ be given,
and let $(S_t)_{t\in\N_0}$ be a random text distributed according to that model.
Then, the (truncated) distribution of accumulated counts
\begin{equation}\label{eqn:dist_of_cdfa_count}
\dist\left(\min\left\{M\,,\,\sum_{i=0}^{\totalsteps-1}\daaemi_{\delta(\state_0,S_0\ldots S_{i})}\right\}\right)
\end{equation}
can be computed in $\OO(\totalsteps\cdot|\stateset|\cdot|\mathcal{C}|^2\cdot|\Sigma|\cdot M)$ time and $\OO(|\stateset|\cdot|\mathcal{C}|\cdot M)$ space. 
If for all $c\in\mathcal{C}$ and $\sigma\in\Sigma$, there exists at most one $c'\in\mathcal{C}$ such that $\varphi(c,\sigma,c')>0$, then the runtime is bounded by $\OO(\totalsteps\cdot |\stateset|\cdot|\mathcal{C}|\cdot |\Sigma| \cdot M)$.
Alternatively, \eqref{eqn:dist_of_cdfa_count} can be computed in $\OO(\log\totalsteps\cdot|\stateset|^3\cdot|\mathcal{C}|^3\cdot M^2)$ time and $\OO(|\stateset|^2\cdot|\mathcal{C}|^2\cdot M)$ space.
\end{theorem}
\begin{proof}
Given the counting DFA $(\stateset,\Sigma,\delta,\state_0,(\daaemi_\state)_{\state\in\stateset})$, we use it to construct the 
DAA $\daa = \big( \stateset, \state_0, \Sigma, \delta, \valset, \val_0, \emiset, (\daaemi_\state)_{\state\in \stateset},(\op_\state)_{\state\in \stateset} \big)$, where
$\valset:=\{0,\ldots,M\}$, $\val_0:=0$, $\emiset:=\{\daaemi_\state:\state\in\stateset\}$, 
and all operations are truncated additions:
\[\op_\state(\val,\emi):=
\begin{cases}
M & \mbox{if }\val+\emi\geq M\,, \\
\val+\emi & \mbox{otherwise}
\end{cases}
\]
for all $\state\in\stateset$. For this DAA~$\daa$, we have
\[  \daaval_{\daa}(s)
  = \min\left\{M\,,\,\sum_{i=0}^{\totalsteps-1}\daaemi_{\delta(\state_0,\prefix{s}{i})}\right\}
\]
for all $s\in\Sigma^*$.

We apply Lemma~\ref{lem:daapaa} to this DAA and text model in order to construct the PAA.
The runtime and space bounds for the basic algorithm follow directly.
To obtain the bounds for the alternative doubling algorithm, namely $\OO(\log\totalsteps\cdot|\stateset|^3\cdot|\mathcal{C}|^3\cdot M^2)$ time and $\OO(|\stateset|^2\cdot|\mathcal{C}|^2\cdot M)$ space,
we exploit that the operations $\op_\state$ are (almost) additions in this case. 
Thus, $U^{(t)}(\state_1,\state_2,\val_1,\val_2)=U^{(t)}(\state_1,\state_2,\val_3,\val_4)$ if $\val_1\leq\val_2<M$, $\val_3\leq \val_4 < M$ and $\val_2-\val_1=\val_4-\val_3$. 
Thus, we can fix $\val_1=0$ and thereby save a factor of $|\valset|=M+1$ in time and space.
The special cases for $\val_2=M$ or $\val_4=M$ can be accommodated in the same bounds.
\end{proof}

Before turning the DFA into a PAA, one may wish to minimize it. Using an algorithm by Hopcroft \cite{Hop71}, a classical DFA can be minimized in $\OO(|\stateset|\log |\stateset|)$ time for an alphabet of constant size, where $\stateset$ is the set of states. Refer to Knuutila \cite{Knu01} for a tutorial-like introduction and a variant that runs in $\OO(|\Sigma|\cdot|\stateset|\log |\stateset|)$ time when the alphabet size is not considered to be a constant. Hopcroft's algorithm can be adapted to minimize counting DFAs by using the partition induced by the different emissions as an initial partition, \ie states with the same emission are grouped together.

In the following, we review some concrete pattern classes important in practice, particularly in computational biology.

\subsubsection{Finite Sets of Strings}\label{sec:aho_corasick}
Assume that the pattern is given in the form of a finite set of strings. In this situation, an Aho-Corasick automaton~\cite{AhoCor75}, which essentially is a DFA, can be built. It can be constructed in linear time by either using the algorithm given in the original paper or by employing a recent elegant algorithm based on the suffix tree of the reverse strings~\cite{DorLan06}. 
The emissions $\daaemi_\state$ (number of matches) can directly be read off the Aho-Corasick automaton's output function for all states~$\state$.

\begin{figure}\centering
\subfigure[PAA for overlapping hits]{\label{fig:model}
\includegraphics[width=.45\textwidth]{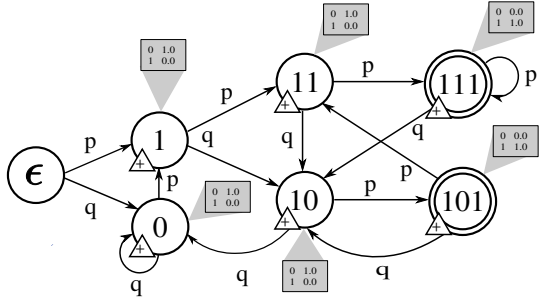}} \hfill
\subfigure[PAA for non-overlapping hits]{\label{fig:model-nov}
\includegraphics[width=.45\textwidth]{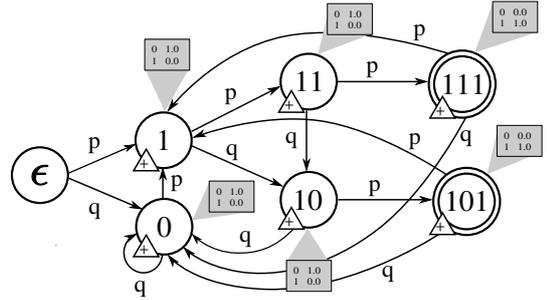}}
\caption{PAAs for the distribution of matches of the pattern set $\{101,111\}$, 
assuming an \iid text model over alphabet $\Sigma=\{0,1\}$ with probability $p$ for character $1$, and $q:=1-p$.
The start state is denoted $\epsilon$. Each state is associated with the operation ``+'' and a Dirac emission distribution,  shown in the gray boxes.}
\label{fig:senspaa}
\end{figure}

Figure~\ref{fig:senspaa} shows a PAA computing the distribution of the number of (a) overlapping matches and (b) non-overlapping matches of the pattern set $\{101,111\}$ in an i.i.d.\ text model over the alphabet $\{0,1\}$, where the probability of seeing $1$ is $p$.

\subsubsection{Finite Sets of Generalized Strings}\label{sec:generalized_strings}
Generalized strings are finite sequences of sets of characters over an alphabet $\Sigma$, for example \texttt{[abc][ac][ab]} (which matches \texttt{aaa}, \texttt{ccb} but not \texttt{aba}). 
Rather than enumerating all strings matching a generalized string, we can directly construct an NFA that recognizes all strings ending with an instance of it. The NFA corresponding to one generalized string is just a linear chain of states; a start state plus one state for each position, where the start state is additionally equipped with a self-transition. The NFA for the set of generalized strings can be constructed by merging all individual start states into one common start state. The next step towards a PAA is to build a DFA. 
To obtain a DFA, we employ the classical subset construction. 
Although, in the worst case, it results in an exponential increase in the number of states, this method is feasible in many practical cases. In fact, the construction procedure can be modified such that it always results in the minimal DFA~\cite{Marschall2010a}.
By the subset construction, a set $B_\state$ of NFA states corresponds to each DFA state $\state$. 
The number of final NFA states in $B_\state$ equals the number of matching generalized strings that end when DFA state~$\state$ is entered, giving us the number of matches $\daaemi_\state$ to be emitted by~$\state$.

\subsubsection*{Prosite Patterns}\label{sec:prosite}
Prosite is a database of biologically meaningful amino acid motifs (see Hulo~et~al.\ \cite{HulBaiBui06}). Prosite patterns can be seen as generalized strings with the extension that, for each position, a ``multiplicity range'' can be specified. In the pattern \texttt{A-x(2,3)-C}, for example, an \texttt{A} is followed by either two or three arbitrary characters followed by a \texttt{C}. We translate every Prosite pattern into a set of generalized strings. The above example would result in the two patterns \texttt{A-x-x-C} and \texttt{A-x-x-x-C}. This set can then be dealt with as explained above.

We implemented the algorithms in Java and ran them on a Intel Core 2 Duo 2.66GHz, 4GB RAM, running Linux to assess practicability. Release 20.17 of Prosite contains 1319 patterns, 16 of which refer to the start or ending of a sequence. Those entries were ignored, leaving a database of 1303 patterns. For 42 patterns (3.2\%) the computation did not succeed due to memory limitations. This can happen if either the Prosite pattern translates into too many generalized strings or if the DFA resulting from the subset construction grows too large. For 1236 of the 1261 remaining patterns, the subset construction was completed within 2 seconds while the computation took 69.9 seconds for the ``worst pattern''. The resulting automata were minimized using Hopcroft's algorithm. Many automata, however, already were minimal or close to minimal; for 1209 automata the minimized automaton was larger than half the size of the original automaton. The majority of resulting minimal automata were of reasonable size: We obtained 1198 automata with less than 10000 states, among which 1036 had less than 500 states.

To give an impression of the runtimes to be expected when computing the distribution of the overlapping occurrence count, consider the pattern 
\[\texttt{C-x-H-R-[GAR]-x(7,8)-[GEKVI]-[NERAQ]-x(4,5)-C-x-[FY]-H}\] 
from the Prosite database. 
It results in an automaton with 462 states. Assuming $M=50$ (maximum number of occurrences of interest) and $\totalsteps=1000$ (text length), computing the distribution of the occurrence count took 1 second.

\subsection{Waiting Time for Pattern Occurrences}
The waiting time for the first occurrence of a pattern equals the waiting time for a state that emits a match. Therefore, its distribution can directly be computed by applying Lemma~\ref{lem:waiting_state}. The waiting time for a subsequent occurrence can be computed by choosing $\alpha$ in Lemma~\ref{lem:waiting_state} to be the equilibrium distribution restricted to all match states, \ie those states that emit a match.

\subsection{Clump Size Distribution}\label{sec:clump_sizes}
We already saw how to use a DFA recognizing a given pattern to construct a PAA. Through this method, we could accurately account for possible self-overlaps of patterns. The structure of self-overlaps was implicitly encoded in the DFA. In this section, we explicitly work out a pattern's tendency to overlap itself by computing its clump size distribution~$\Psi$. As detailed in~\cite{Marschall2009}, the exact clump size distribution of a pattern is, besides its theoretical value, useful for the construction of compound Poisson approximations. Compound Poisson approximations have also been discussed in~\cite{Schbath1995,Waterman1995,Roquain2007}.

\begin{definition}
Given a sequence~$s\in\Sigma^\ast$ and a pattern~$\pattern$, a \emph{clump} is a maximal set of overlapping occurrences of~$\pattern$ in~$s$.
\end{definition}

For example, let $\pattern:=\texttt{ACA}$ and $s:=\texttt{G\underline{ACACA}TT\underline{ACA}AA}$. Then $s$ contains three occurrences of $\pattern$ in two clumps (underlined).
By definition, a clump consists of at least one match. We call the position of match's last character \emph{match position} and consider the first match position in a clump. Further, we call the distribution of PAA states at such positions \emph{clump start distribution} and denote it $\gamma$; \ie given that $j$ is the first match position in a clump, then $\prob(\stateproc_j\!=\!\state) =: \gamma(\state)$, which is asymptotically independent of $j$ under certain assumptions.
For now, we assume $\gamma$ to be known and come back to the task of its calculation later.

\begin{figure}
\begin{center}
\includegraphics[width=.7\textwidth]{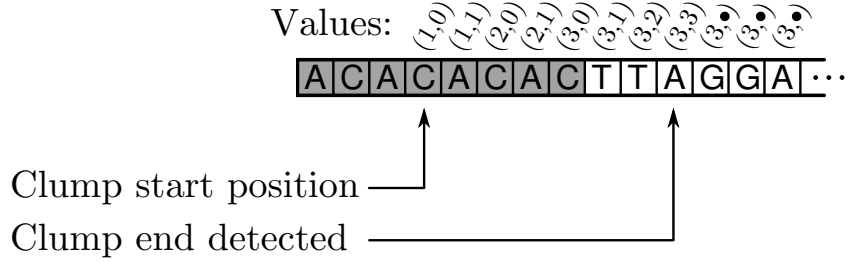}
\end{center}
\caption{A clump (shaded gray) of three occurrences of the pattern \texttt{ACAC}.
By definition, the clump starts at the last character of the first occurrence. 
The values computed by the clump size DAA are shown above the string. 
As soon as the second counter reaches $\patlen-1=3$, the clump has ended.}
\label{fig:clump}
\end{figure}

If $\patlen\geq 2$ is the length of the given motif, then a clump ends if $\patlen-1$ consecutively visited states do not emit a match. That means we need to keep track of (a)~the number of non-match states consecutively visited and (b)~the number of matches the clump contains so far. 
The PAA framework allows this by modifying the PAAs described in the previous subsections. 
We define a new value set $\valset':=\N_0\times\{0,\ldots,\patlen-1,\bullet\}$ with the start value $\val_0':=(0,0)$ and attach the following semantics: If we are in state $\state$ and the current value is $(h,x)$, we have seen $h$ matches in the current clump and the last of these matches occurred $x$ steps in the past; \ie if $x=0$, a match has been emitted from the current state. The special value $x=\bullet$ indicates that the clump has ended. We define the operations accordingly:
\[\op_\state': \big((h,x),\emi\big)\mapsto
\begin{cases}
(h+\emi,0) & \mbox{if $\emi>0$ and $x\in\{0,\ldots,\patlen-2\}$\,,} \\
(h,x+1) & \mbox{if $\emi=0$ and $x\in\{0,\ldots,\patlen-2\}$\,,} \\
(h,\bullet) & \mbox{if $x\in\{\patlen-1,\bullet\}$\,.}
\end{cases}
\]
In other words, if a match has been found ($\emi>0$), we increase the number of matches~$h$ by~$\emi$ and reset the distance to the last match to 0. Otherwise ($\emi=0$, no match occurred), $h$ remains unmodified, but the number of steps~$x$ since the last match is increased by one. See Figure~\ref{fig:clump} for an example.
To incorporate the clump start distribution~$\gamma$, we use one additional state $\state_0'$ that becomes the new start state; consequently, we set $\stateset':=\lbrace \state_0'\rbrace\cup \stateset$ and define the new transition function to be
\begin{equation}\label{eqn:clump_size_paa_function}
\paatrans': (\state,\state')\mapsto
\begin{cases}
  \gamma(\state') & \mbox{if $\state=\state_0'$\,,} \\
  \paatrans(\state,\state')     & \mbox{otherwise\,.}
\end{cases}
\end{equation}
The set $\valset'$ is infinite. As discussed in Section~\ref{sec:statevalue}, this does not pose a problem as the range of each $\valproc_t$ is finite. Furthermore, for many applications it is sufficient to truncate the clump size distribution and use the value set $\valset'':=\lbrace 1,\ldots,M\rbrace\times\lbrace 0,\ldots,\patlen-1,\bullet\rbrace$ along with adapted operations $\op_\state''$. Employing one of the algorithms shown in Sections~\ref{sec:evo_basic} and~\ref{sec:evo_doubling}, respectively, we can then calculate the joint state-value distributions $\rho_t(\state,h,x):=\prob\big(\stateproc_t=\state,\valproc_t=(h,x)\big)$.
A clump ends if no new match has occurred $\patlen-1$ steps after the previous match.
The clump size distribution $\Psi$ is thus given by
\begin{equation}\label{eqn:clump_len_dist}
\Psi(h)=\sum_{t=0}^\infty\sum_{\state\in \stateset}\rho_t(\state,h,\patlen-1)\mbox{\,.}
\end{equation}
To actually compute $\Psi$, we start with the initial table $\rho_0$ and iteratively calculate the tables $\rho_t$ for larger~$t$. Each $\rho_t$ contributes to the sought distribution through the inner sum from Equation~\eqref{eqn:clump_len_dist} and we can successively add the contributions to an intermediate clump size distribution. Observe that the difference between the intermediate clump size distribution after iteration~$t$ and the exact one is bounded by
\[1-\sum_{\state\in\stateset}\sum_{h=0}^M\rho_t(q,h,\bullet)\,.\]
Thus, we iterate until this quantity drops under an accuracy threshold. The number of necessary steps, however, is bounded by $\OO(M\cdot \patlen)$, because a clump containing $M$ matches can have a length of at most $\OO(M\cdot \patlen)$. In total, we need $\OO(|\Sigma|\cdot|\mathcal{C}|\cdot|\stateset|\cdot M^2\cdot \patlen^3)$ time to compute the exact clump size distribution. Again, a factor of~$|\mathcal{C}|$ can be saved if for all $c\in\mathcal{C}$ and $\sigma\in\Sigma$, there exists at most one $c'\in\mathcal{C}$ such that $\varphi(c,\sigma,c')>0$.

\subsubsection*{State Distribution at Clump Start}
Let us come back to computing the clump start distribution $\gamma$ needed in Equation~\eqref{eqn:clump_size_paa_function}. 
As discussed in Section~\ref{sec:paa}, the PAA's state process $(\stateproc_t)_{t\in\N_0}$ is a Markov chain and, hence, the classical theorems about existence of and convergence to an equilibrium distribution apply: Irreducibility and aperiodicity are sufficient for convergence to a unique equilibrium distribution. Assuming (a) that a pattern does not start with a wildcard and (b) all text model states $c\in\mathcal{C}$ have positive occurrence probability, these conditions can be verified to be fulfilled by construction of the PAA.

We consider the joint distribution of state and steps since the last match position. 
Recall that $\emiproc_t$ is the emission process of a PAA. 
We define $L_t$ as the number of steps since we last encountered a match before step $t$. Thus
\[L_t := \min \big\{t' \in \{1,\dots,t\} \,\big|\, \emiproc_{t-t'}>0\big\},\]
where we set $L_t := -\infty$ if the set is empty, meaning that no match has occurred until step~$t-1$.
Again we use the PAA framework to compute the joint state-value distribution $\dist(\stateproc_t,L_t)$ for any desired $t$. The clump start distribution is now given by
\begin{equation}
\label{eqn:limit_clump_start_dist}
\gamma(\state) = \lim_{t\rightarrow\infty}\prob\big(\stateproc_t\!=\!\state\,\big|\,L_t\geq \patlen\,,\emiproc_t>0\big)\mbox{\,.}
\end{equation}
In practice, the limits for $t\to\infty$ exist and converge in a few steps to double precision.


\section{Analysis of Window-Based Pattern Matching Algorithms}\label{sec:horspool}
The basic pattern matching problem is to find all occurrences of a \emph{pattern} string in a (long) \emph{text} string as fast as possible.
Let~$\totalsteps$ be the text length and~$\patlen$ be the pattern length.
The well-known Knuth-Morris-Pratt algorithm~\cite{KnuthMorrisPratt1977} reads each text character exactly once from left to right after preprocessing the pattern and needs a total of~$\Theta(\totalsteps+\patlen)$ character accesses. 
In contrast, the Boyer-Moore~\cite{Boyer1977}, Horspool~\cite{Horspool1980} and Sunday~\cite{Sunday1990} algorithms move a length-$\patlen$ search window across the text and first compare its \emph{last} character to the last character of the pattern.
This often allows to move the search window by more than one position (at best, by~$\patlen$ positions if the last window character does not occur in the pattern at all), for a best case of~$\Theta(\patlen+\totalsteps/\patlen)$ but a worst case of~$\Theta(\patlen+\patlen\totalsteps)$ character accesses. 
The worst case can be improved to~$\Theta(\patlen+\totalsteps)$, but this makes the code more complicated and is seldom useful in practice.
The Horspool algorithm and the variant of Sunday can be seen as modifications of the Boyer-Moore algorithm that are simpler to implement and additionally perform better in practice~\cite{NavRaf02}. In general, a window-based algorithm that searches for a pattern~$\pattern$ is characterized by
\begin{itemize}
 \item a window size $z$,
 \item a cost function $\windowcost{\pattern}:\Sigma^z\to\N_0$ giving the cost caused by a window,
 \item a shift function $\shift^\pattern:\Sigma^z\to\{1,\ldots,\patlen\}$ giving the number of positions the window can safely be shifted by.
\end{itemize}
The total cost of processing a text $s\in\Sigma^\totalsteps$ is denoted $\cost{\pattern}(s)$.
In this section, we develop a methodology to calculate the exact distribution of $\cost{\pattern}(S_0\ldots S_{\totalsteps-1})$ when $S$ is a random text and pattern $\pattern$ is fixed. This question has so far not been investigated, even though related questions have been answered in the literature.
For example, \cite{Baeza-Yates1990,BaezaYatesRegnier1992} analyze the expected number of character accesses for Horspool's algorithm. In~\cite{Mahmoud1997} it is shown that the number of character accesses is asymptotically normally distributed for i.i.d.\ texts, and~\cite{Smythe2001} extends this result to Markovian text models.

\begin{algorithm}[t!]
\caption{\label{alg:horspool}\textsc{Horspool}}
\begin{algorithmic}[1]
\vspace*{.1cm}
\REQUIRE text $s\in\Sigma^\ast$, pattern $\pattern\in\Sigma^\patlen$
\ENSURE pair (number $occ$ of occurrences of $\pattern$ in $s$, number $cost$ of accesses to $s$)
  \STATE pre-compute table $\shift[\sigma]$ for all $\sigma\in\Sigma$ 
	\STATE $(occ,cost) \gets (0,0)$
 	\STATE $t \gets \patlen-1$
	\WHILE{$t < \len{s}$}\label{alg:hor_mainloop_begin}
		\STATE $i\gets 0$
		\WHILE{$i<\patlen$}\label{alg:hor_innerloop_begin}
			\STATE $cost\gets cost+1$
			\STATE \textbf{if} $\chr{s}{t-i}\neq\chr{\pattern}{(\patlen-1)-i}$ \textbf{then break}
			\STATE $i\gets i+1$
		\ENDWHILE\label{alg:hor_innerloop_end}
		\STATE \textbf{if} $i=\patlen$ \textbf{then} $occ\gets occ+1$
		\STATE $t\gets t+\shift[\chr{s}{t}]$
	\ENDWHILE\label{alg:hor_mainloop_end}
	\RETURN $(occ, cost)$\label{alg:hor_n_alteration}
\end{algorithmic}
\end{algorithm}

The technique we introduce here allows to compute the exact distribution of the total cost for arbitrary cost functions, general finite-memory text models as defined in Section~\ref{sec:paa_on_randseqs}, and applies to all window-based pattern matching algorithms. For concreteness, we consider Horspool's algorithm as given in Algorithm~\ref{alg:horspool}, that means, pattern and window are compared from right to left and the shift solely depends on the window's last character. Formally, Horspool's algorithm is characterized by
\begin{itemize}
 \item $z:=\patlen$ (the window size equals the pattern length),
 \item $
   \windowcost{\pattern}(w) := \begin{cases}
   \patlen &\text{ if } \pattern=w,\\
   \min\big\{i: 1\leq i\leq \patlen,\; \chr{\pattern}{\patlen-i} \neq \chr{w}{\patlen-i}\big\} &\text{ otherwise},
   \end{cases}$
 \item the shift depends on the position of the rightmost occurrence of~$\chr{w}{\patlen-1}$ in~$\pattern$:
 \begin{align*}
   \rightpos^\pattern(w) &:= \max\big[ \{i\in\{0,\dots,\patlen-2\}: \chr{\pattern}{i}=\chr{w}{\patlen-1}\} \cup \{-1\} \big]\,,\\
   \shift^\pattern(w) &:= (\patlen-1) - \rightpos^\pattern(w)\,.
\end{align*}
\end{itemize}
Here, we have used the number of character accesses as a measure of cost.

We now construct a PAA that allows to compute the cost distribution for arbitrary window-based pattern matching algorithms with respect to a random text of length~$\totalsteps$ defined by a text model $(\mathcal{C},c_0,\Sigma,\varphi)$. 
Note that we cannot construct a DAA and apply Lemma~\ref{lem:daapaa}, as the considered algorithms can read several characters ``at once'', while a DAA is only capable of processing one character at a time.
In the defined model of window-based algorithms, shift and cost depend solely on the current window. 
Therefore, we model each possible window as a state (although concrete algorithms may permit smaller state spaces). 
Additionally, we need to keep track of the current text model state after generating the current window. 
We use the state space 
\[\stateset:=\{(\emptystring,c_0)\}\cup\left(\Sigma^z\times\mathcal{C}\right)\,,\]
where $(\emptystring,c_0)=:\state_0$ is the start state and any other state~$(w,c)$ corresponds to window content~$w$ and text model state~$c$. Note that, if the text model is a Markovian model of order $r\leq z$, then its state is fully determined by the current window. That means, only $|\Sigma|^z+1$ different states are reachable, all others can be discarded. For arbitrary text models, however, more states might be reachable.
The strategy to construct the PAA is to simulate the algorithm and count both the reached position in the text and the cost so far. To this end, we assume that the text length $\totalsteps$ is fixed and define the value set
\[\valset := \{0,\ldots,\totalsteps-1,\bullet\} \times \{0,\dots,\totalsteps\cdot\maxwindowcost{\pattern}\}\,,\]
where $\maxwindowcost{\pattern}:=\max\{\windowcost{\pattern}(w):w\in\Sigma^z\}$,
with $\maxwindowcost{\pattern}=m$ for Horspool's algorithm.
Value $\val=(t,\xi)$ corresponds to the current window ending at text position $t$ (with $\bullet$ indicating that the text has ended), having accumulated a cost of~$\xi$ so far. The start value is set to~$\val_0:=(z-1,0)$.

Each state deterministically emits cost and shift of the associated window, therefore we set \[\emiset:=\{1,\ldots,\patlen\}\times\{0,\ldots,\maxwindowcost{\pattern}\}\,,\]
where an emission of $(t',\xi')\in\emiset$ indicates that the current window has caused a cost of~$\xi'$ and the window is to be shifted by~$t'$. Formally, we define
\[\emidist_{(w,c)}\big((t',\xi')\big):=
\begin{cases}
1 & \mbox{if }t'=\shift^\pattern(w)\mbox{ and }\xi'=\windowcost{\pattern}(w)\,,\\
0 & \mbox{otherwise}\,,
\end{cases}\]
for all $(w,c)\in\stateset$. Note that we could use other (non-Dirac) emission distributions if necessary. We might assume, for example, that a cache miss occurs with a certain probability and associate higher costs with this event.
The operation $\op_\state$ in each state is essentially an addition on $\valset$ with one exception: To indicate that the complete text has been processed, we use the special values $(\bullet,\xi)\in\valset$. We define:
\[\op_\state\big((t,\xi),(t',\xi')\big):=
\begin{cases}
(t+t',\xi+\xi') & \mbox{if }t\neq\bullet\mbox{ and }t+t'<\totalsteps\,, \\
(\bullet,\xi+\xi') & \mbox{if }t\neq\bullet\mbox{ and }t+t'\geq\totalsteps\,, \\
(\bullet,\xi) & \mbox{otherwise}\,.
\end{cases}\]
The last component to be specified is the transition function~$\paatrans$. The shift, and therefore all possible target states, are determined by the current window contents, that means, the suffix of the current window must match the prefix of the subsequent window. We define an indicator function that tells whether two windows are compatible in this sense:
\[I(w,w'):=\big\iverl\suffix{w}{\shift^\pattern(w)}=\prefix{w'}{z-1-\shift^\pattern(w)}\big\iverr\,.\]
While $I(w,w')$ tells whether a transition from~$w$ to~$w'$ is allowed, its probability is given by the text model.
\[T\big((w,c),(w',c')\big):=
\begin{cases}
\prob\left(c_0\xrightarrow{w}c'\right) & \mbox{if }(w,c)=(\emptystring,c_0)\,, \\
I(w,w')\cdot\prob\left(c\xrightarrow{\suffix{w'}{z-\shift^\pattern(w)}}c'\right)& \mbox{otherwise}\,,
\end{cases}\]
where 
\[\prob\left(c_0'\xrightarrow{\sigma_0\ldots\sigma_{j-1}}c_j'\right):=\sum_{c_1',\ldots,c_{j-1}'\in\mathcal{C}}\ \prod_{i=0}^{j-1}\varphi(c_i',\sigma_i,c_{i+1}')\]
is the probability that the text model $(\mathcal{C},c_0,\Sigma,\varphi)$ produces the string $\sigma[0]\ldots\sigma[j-1]\in\Sigma^j$ while going from state~$c_0'\in\mathcal{C}$ to state~$c_j'\in\mathcal{C}$ in~$j$ steps.

\begin{figure}[t!]\centering
\includegraphics[width=\textwidth]{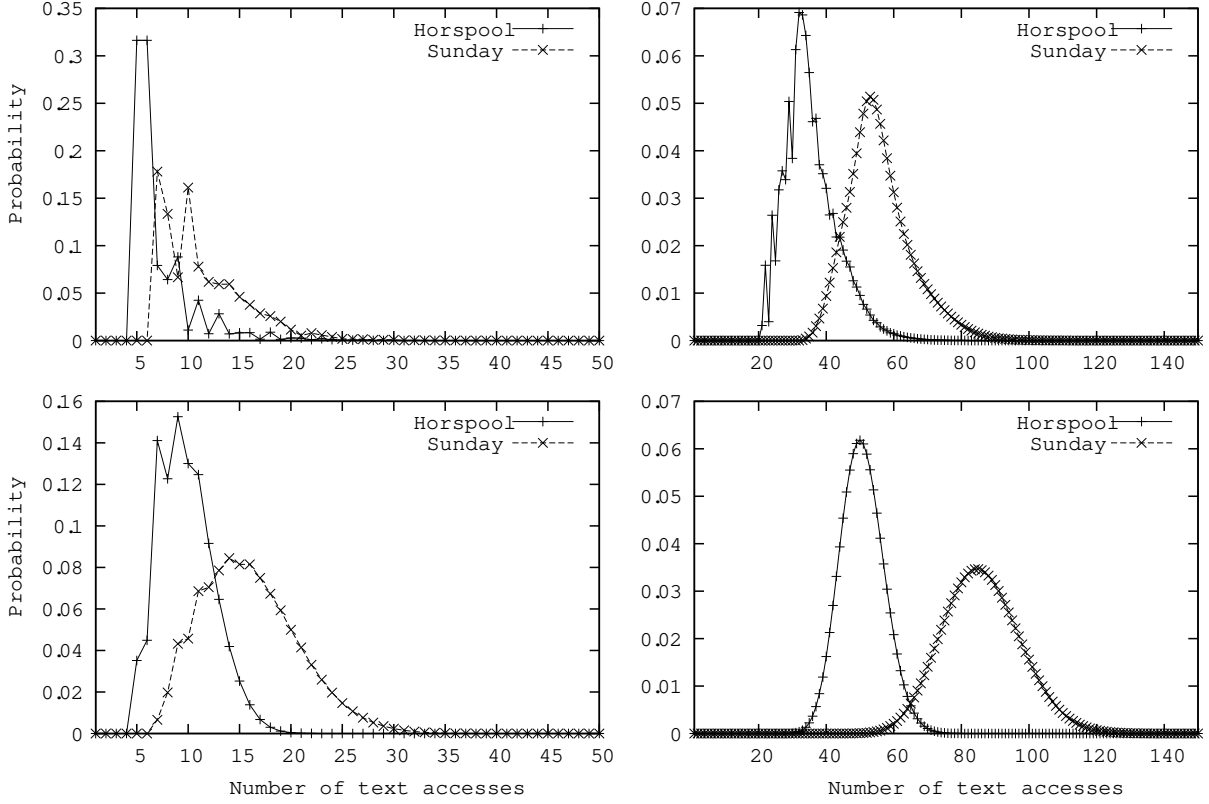}
\caption{Exact distribution of the number of text accesses for Horspool's and Sunday's algorithms using a uniform i.i.d.\ text model over the alphabet $\{\texttt{A},\texttt{C},\texttt{G},\texttt{T}\}$. Top vs.\ bottom row: pattern \texttt{AAAAA} vs.\ \texttt{ACAGC}. Left vs.\ right column: text length 20 vs.\ 100.
}
\label{fig:horspool_distributions}
\end{figure}

\begin{theorem}
Let a window-based pattern matching algorithm specified by $z$, $\windowcost{\pattern}$, and $\shift^\pattern$ and a text model $(\mathcal{C},c_0,\Sigma,\varphi)$ be given. 
The exact distribution of the cost of processing a random string of length~$\totalsteps$ can be computed in $\OO(\totalsteps^3\cdot|\Sigma|^{2z}\cdot|\mathcal{C}|\cdot\maxwindowcost{\pattern})$ time and $\OO(\totalsteps^2\cdot|\Sigma|^z\cdot|\mathcal{C}|\cdot\maxwindowcost{\pattern})$ space. If the text model is Markovian of order $r\leq z$, then $\OO(\totalsteps^3\cdot|\Sigma|^{2z}\cdot\maxwindowcost{\pattern})$ time and $\OO(\totalsteps^2\cdot|\Sigma|^z\cdot\maxwindowcost{\pattern})$ space are sufficient.
\end{theorem}
\begin{proof}
By construction, the PAA simulates the working of the specified algorithm. Assuming $|\emiset|=\OO(1)$, runtime and space bounds follow from Lemma~\ref{lem:paa_basic}. If the text model is Markovian of order $r\leq z$, then only $|\Sigma|^z+1$ different states are reachable. The claimed time and space bound can then be met by using a reduced state space $\stateset':=\{\emptystring\}\cup|\Sigma|^z$.
\end{proof}

The exponential dependency on the window length~$z$ allows practical computations only for short patterns. For a pattern length of 5 and text lengths 20 and 100, a comparison of Horspool's and Sunday's algorithms is shown in Figure~\ref{fig:horspool_distributions}. The calculation took 1.8~seconds and 40.4~seconds for text length 20 and 100, respectively\footnote{See \texttt{http://www.rahmannlab.de/software} for an implementation in JAVA. The experiments were run on an Intel Core 2 Quad CPU at 2.66GHz.}. The plots reveal the combinatorial nature of the number of text accesses (which is asymptotically normally distributed) for short texts. It also reveals that Sunday's algorithms needs more character accesses than Horspool's algorithm for the example patterns.

\section{Alignment Seed Sensitivity}\label{sec:seeds}

We describe an application of the PAA framework to determine the quality of seeds used in homology search. 
Homologous biosequences have developed from a common ancestor and usually share high sequence similarity.
In homology search, a sequence database is searched for a query sequence in order to find potential homologs, \ie evolutionarily related sequences. 
To this end, each database sequence is compared with the query, which can be done by a local alignment method such as the Smith-Waterman algorithm~\cite{Waterman81}.

\subsection{Related Work and New Results} 
Since exact local alignment is too slow in practice, most heuristic homology search algorithms are based on a two-phase filtration technique \cite{Pearson1988,Altschul90,Altschul1997,Kent2002}. 
First, candidate sequences are selected that share a common pattern (``seed'') of matching characters with the query.
These candidates (or ``hits'') are then further investigated by an exact method. 
Initially, \emph{contiguous seeds} (e.g., perfectly  matching DNA 11-mers in the initial BLAST implementation) were used.
PatternHunter (PH) by Ma~et~al.\ \cite{Ma2002} was the first tool to systematically advocate and investigate \emph{spaced seeds}:
PH looks for 18-mers with at least 11 matching positions distributed as \texttt{111*1**1*1**11*111}, where \texttt{1} denotes a necessary match and \texttt{*} denotes a don't care position (match or mismatch). 
Over time, various seed models have been proposed in the literature, including consecutive seeds~\cite{Pearson1988,Altschul90}, spaced seeds~\cite{Ma2002,Buhler2003,Brejova2004,Choi2004}, subset seeds~\cite{Kucherov2006}, vector seeds~\cite{Brejova2005}, and indel seeds~\cite{Mak2006}. 

In the context of homology search, it is customary to model random \emph{alignments} instead of random sequences.
Typical models for such alignments may consist of several homology parameters an hence called \emph{homology model}s.
They are described in Section~\ref{sec:homologymodel}.
Different seeds in a class (e.g.\ all seeds with 11 match positions and length 18) can be compared according to their \emph{sensitivity}, \ie the probability to ``hit'' a random alignment of given length from a given homology model (see Definition~\ref{def:hit} in Section~\ref{sec:seedmodel} for a formal definition of ``hit'').

A good seed exhibits high sensitivity for alignments that model evolutionarily related biosequences, and low sensitivity values for alignments that represent unrelated sequences. 
The latter property ensures that the seed does not detect too many \emph{random hits}. 
Random hits decrease the efficiency of the filtration phase, since they are checked in vain for a significant alignment. 
An interesting finding was that the PH approach led to an increase in both sensitivity and filtration efficiency, compared to seeds of contiguous matches.
Based on the observations in \cite{Ma2002}, the advantages of spaced seeds over consecutive seeds have been subsequently evaluated by many authors \cite{Buhler2003,ChoiZhang2004,Li2006}. 

An extension to single seed models is the design of a multiple seed. 
This is a set of spaced seeds to be used simultaneously, such that a similarity is detected when it is found by (at least) one of the seeds. 
The idea to use a family of spaced seeds for BLAST-type DNA local alignment has been suggested by Ma~et~al.\ \cite{Ma2002} and was implemented in PatternHunter~II \cite{Li2004}. 
It has also been applied to local protein alignment in \cite{Brown2005}. 
Recent approaches \cite{Kong2007,Ilie2007} approximate the sensitivity of multiple spaced seeds by means of correlation functions. 
Since finding optimal multiple seeds is challenging, most authors concentrate on the design of efficient sets of seeds, leading to higher sensitivity than optimal single seeds \cite{Li2004,Kucherov2005,Sun2005}. 

When searching optimal seeds, one faces the following problems to evaluate candidate seeds:
\begin{problem}[Sensitivity computation]\label{prob1}
Given a homology model, a target length~$\totalsteps$, and a set of seeds, 
what is the probability that a random alignment of length~$\totalsteps$ is hit by the seed (at least once)?
\end{problem}
\begin{problem}[Hit distribution]\label{prob2}
Given a homology model, a target length~$\totalsteps$, a set of seeds, and a maximal hit number $K$,
what is the probability that a random alignment of length~$\totalsteps$ is hit by the seed exactly $k$ times (or at least $k$ times), for each $k=0,\dots, K$,
when counting (a) overlapping hits, (b) non-overlapping hits?
\end{problem}

The second, more general question has not yet been investigated in the literature.
As we show, the distribution is directly provided by the constructed PAA and allows the investigation of optimality criteria different from sensitivity alone.

\subsection{Homology Models}\label{sec:homologymodel}
We describe random alignments with known degree of similarity (e.g.\ a certain per cent identity value) by means of a \emph{homology model}. 
A homology model generates representative strings $\mc{A}$ over an alphabet $\Sigma$ indicating the status of the alignment columns. 
In the simplest and most frequently studied case only substitution mutations and ungapped alignments are considered \cite{Ma2002,Buhler2003,Brejova2004,Choi2004,ChoiZhang2004,Sun2005}; see Table~\ref{tab:alignmentstring}.
That is $\Sigma=\{0,1\}$, referring to matches ($1$) and mismatches ($0$). 
\begin{table}[tb]\centering
\caption{\label{tab:alignmentstring}Representative string $\mc{A}$ of an ungapped alignment between two sequences.}
  \begin{tabular}{l|lllllllllll}
  Query    & G&C&G&A&A&T&G&C&C&T\\
  Database & G&C&C&A&A&C&G&C&T&T\\\hline
  $\mc{A}$ & $1$&$1$&$0$&$1$&$1$&$0$&$1$&$1$&$0$&$1$
  \end{tabular}
\end{table}

Indel seeds, designed for gapped alignments, use the alignment alphabet $\Sigma=\{0,1,2,3\}$, where additionally $2$ denotes an insertion into the database sequence, and $3$ indicates an insertion into the query sequence. 
There are various other alignment alphabets, e.g.\ the ternary alphabet representing a match or transition or transversion in DNA \cite{Noe2004}, or even larger alphabets to distinguish different pairs of amino acids in the case of proteins \cite{Brown2005}.
A representative string is modeled as a Markov chain $(\Sigma,P,p^{0})$ with a transition matrix $P$ and an initial distribution $p^{0}$ on the alphabet $\Sigma$.
$P$ and $p^{0}$ are called \emph{homology parameters}.

\paragraph{Ungapped alignments.}
We model ungapped alignments by an \iid homology model with $\Sigma=\{0,1\}$.
In this case, the transition probability $P(\sigma,\sigma')$ does not depend on $\sigma$.
In particular, $P$ takes the form 
$\begin{pmatrix}1-p && p \\ 1-p && p\end{pmatrix}$
for a \emph{match probability} $p\in[0,1]$, which quantifies the average identity of such alignments;
for example, $p\approx 0.3$ for unrelated, $p\approx 0.95$ for closely related DNA sequences.

\paragraph{Gapped alignments.}
For gapped alignments, we use a first-order Markov chain.
This is appropriate since the respective homology model should prohibit the pairs `$23$' and `$32$' in a representative string, because a substitution is more plausible than two consecutive indels.
For our calculations, we used the transition matrix $P$ proposed in \cite{Mak2006}:
\begin{equation}\label{indelMat}
\bordermatrix{
&0&1&2&3\cr
0&p_{0}&p_{1}&p_{\text{g}}&p_{\text{g}}\cr
1&p_{0}&p_{1}&p_{\text{g}}&p_{\text{g}}\cr
2&p^{*}_{0}&p^{*}_{1}&p_{\text{g}}&0\cr
3&p^{*}_{0}&p^{*}_{1}&0&p_{\text{g}}\cr
},
\end{equation}
where $p_{0}$ is the probability of a mismatch, $p_{1}$ is the probability of a match, and $p_{\text{g}}$ refers to the probability of a gap in the alignment.
In order to obtain a stochastic transition matrix, $p^{*}_{\sigma} = p_{\sigma} + p_{\text{g}} p_{\sigma} / (p_{0}+p_{1})$ for $\sigma\in\{0,1\}$ redistributes $p_{g}$ to match and mismatch characters.
The initial distribution is given by $p^{0} = (p_{0},p_{1},p_{\text{g}},p_{\text{g}})$.
Other transition probabilities are possible, e.g.\ if alignments with affine gap costs should be modeled.

\subsection{Seed Models}\label{sec:seedmodel}
A seed $\pi=\pi[0]\pi[1]\dots \pi[L-1]$ is a string over an alphabet of ``care'' and ``don't care'' characters.
It represents alignment regions that indicate matches at the ``care'' positions.
A seed is classified $(L,\omega)$ by its \emph{length} $L=|\pi|$ and its \emph{weight} $\omega$, which refers to the number of ``care'' positions.

A \emph{contiguous seed} represents a region of contiguous matches, \ie $\pi[i]=\texttt{1}$ for $0\leq i < L$. 
A \emph{spaced seed} is a string over the alphabet $\Xi=\{\texttt{1},\texttt{*}\}$. 
The ``care'' positions are indicated by \texttt{1}, while \texttt{*} refers to a match/mismatch wildcard. 
Reasonable seeds for the purpose of homology search always require $\pi[0]=\pi[L-1]=\texttt{1}$.
An \emph{indel seed} according to Mak~et~al.\ \cite{Mak2006} is a string over the alphabet $\Xi=\{\texttt{1},\texttt{*}, \texttt{?}\}$, where \texttt{1} and \texttt{*} are as above, and \texttt{?} stands for zero or one character from the alignment alphabet $\Sigma=\{0,1,2,3\}$. 
Two consecutive \texttt{?} symbols represent any character pair except `$23$' or `$32$'.
By means of this interpretation, the model explicitly allows for indels of variable size. 
For example, \texttt{1??1} may detect indels of size $0$, $1$, or $2$. 
It hence tolerates $2$, $1$, or $0$ match/mismatch positions. 

A seed can thus be converted to a generalized string (see Section~\ref{sec:generalized_strings}) over the alignment alphabet when we additionally allow to skip some characters ($\epsilon$-characters).
In any case, a seed can be represented as a finite set of patterns over the alignment alphabet;
this pattern set $\ps(\pi)$ contains all instances of the generalized string.
\begin{definition}[Hit]\label{def:hit}
A \emph{hit} of a seed $\pi$ in an alignment $\mc{A}$ is an occurrence\footnote{A hit has previously been called a match or an occurrence, but here a match concerns a single position in an alignment, and the term ``hit'' seems more descriptive.} of a string from $\ps(\pi)$ in $\mc{A}$.
We call an ending position of a seed hit in $\mc{A}$ a \emph{hit position}.
\end{definition}

\begin{example}
Consider the indel seed $\pi= \texttt{1*1?1}$. It corresponds to the generalized string $[1][01][1][\epsilon 0123][1]$, and the instances are given by the pattern set 
\[\ps(\pi)=\{1011,1111,\\10101,10111,10121,10131,11101,11111,11121,11131\}.\].
\noindent The seed hits the alignment string $1011011110$ at positions $3$, $6$, $7$, and $8$, respectively.
\end{example}

For a finite, non-empty set $\Pi=\{\pi_{1},\dots ,\pi_{m}\}$ of spaced seeds, also called \emph{multiple spaced seed}, 
the patterns are collected in $\mc{PS}(\Pi)=\cup_{i=1}^{m}\mc{PS}(\pi_{i})$. 
A multiple seed is said to \emph{hit} $\mc{A}$, if at least one of its components does.

\subsection{PAAs for Seed Sensitivity and Applications}\label{subsec:paaseed}   
With the seed (set) $\Pi$ represented as a pattern set $\mc{PS}(\Pi)$, and the alignment model being a finite memory text model (see Section~\ref{sec:text_model}), the problem has been reduced to computing the hit distribution of a finite set of patterns (Section~\ref{sec:aho_corasick}).
As mentioned in Section~\ref{subsec:dfa2daa}, both overlapping and non-overlapping seed hits can be considered.
In fact, Figure~\ref{fig:senspaa} in Section~\ref{sec:aho_corasick} shows the PAA for seed $\pi=\texttt{1*1}$ with pattern set $\mc{PS}(\pi)=\{111,101\}$ in an \iid text model to count (a) overlapping and (b) non-overlapping hits.

\begin{table}[t!]\centering
 \caption{\label{tab:seeds}Comparison of a seed requiring 11 contiguous matches in an alignment and the PH seed requiring at least 11 matches within 18 alignment columns at particular positions.
 The alignment model is ungapped i.i.d., with $p$ being the probability of a match, and $1-p$ being the probability of a mismatch. We consider both highly similar sequences (top, $p=0.95$) and unrelated sequences (bottom, $p=0.3$). The tables show the probabilities for exactly $k$ overlapping hits, for $k=0,\dots,3$, in a target region of length 64.}
 \begin{tabular}{l|llll}
  $p=0.95$                     & $k=0$ & $k=1$ & $k=2$ & $k=3$\\\hline
  \texttt{11111111111}         & $4.1285\cdot 10^{-4}$ & $5.4005\cdot 10^{-4}$ & $8.4467\cdot 10^{-4}$ & $0.0012$ \\
  \texttt{111*1**1*1**11*111}  & $6.7331\cdot 10^{-6}$ & $4.4978\cdot 10^{-5}$ & $1.6120\cdot 10^{-4}$ & $4.1669\cdot 10^{-4}$\\\hline\hline
  $p=0.3$                      & $k=0$ & $k=1$ & $k=2$ & $k=3$\\\hline
  \texttt{11111111111}         & $0.9999325$ & $4.7615\cdot 10^{-5}$ & $1.4025\cdot 10^{-5}$ & $4.1295\cdot 10^{-6}$ \\
  \texttt{111*1**1*1**11*111}  & $0.9999170$ & $8.2780\cdot 10^{-5}$ & $2.2438\cdot 10^{-7}$ & $8.9947\cdot 10^{-9}$ \\
 \end{tabular}
\end{table}

In contrast to previous work that only considers the sensitivity (probability of at least one hit, Problem~\ref{prob1}),
the PAA framework yields the entire match distribution (Problem~\ref{prob2}).
However, if only the sensitivity is desired, the value set can be reduced to $\valset=\{0,1\}$,
reducing requirements to $\OO(\totalsteps|\stateset|^2)$ time and $\OO(|\stateset|)$ space.

An example comparing a contiguous seed with the PH seed is shown in Table~\ref{tab:seeds}.
For match probability $p=0.95$ (an alignment of highly similar sequences), a good seed should achieve a high sensitvity, or low probability for zero hits; indeed, the PH seed is almost two orders of magnitude better than the contiguous seed ($6.7\cdot 10^{-6}$ vs.\ $4.1 \cdot 10^{-4}$ for $k=0$ hits).
For $p=0.3$ (essentially a random alignment), a good seed should achieve a low sensitivity.
Both sensitivty values are comparable ($6.75\cdot 10^{-5}$ for the contiguous seed; $8.3\cdot 10^{-5}$ for the PH seed).
If we consider only candidates with at least two hits, the PH seed retains its advantage for alignments of highly similar sequences and now outperforms the contiguous seed even for random alignments (probability of $3.5\cdot 10^{-5}$ for at least two matches for the contiguous seed, but only $2.2 \cdot 10^{-7}$ for the PH seed).

The PAA framework provides a unifying method for computing seed sensitivity for different alignment models and seed models that were so far developed in an \textit{ad-hoc} fashion in the literature.
Furthermore, it is easily extended; let us mention two examples.

For a simple homology model with few parameters (say, the \iid homology model for ungapped alignments with a single parameter $p$), it is possible to evaluate the recurrence~\ref{eqn:basic_recurrence} symbolically, i.e., by representing the entries of the transition matrix $\paatrans(\state',\state)$ and the probabilities of the state-value distribution $f_t(\state,\val)$ as polynomials in the parameters.
The resulting polynomial only needs to be computed once; then one can assess the sensitivity of a seed under different parameter values. This was previously presented by Mak~et~al.\ \cite{Mak2007}, without using the PAA approach.

The PAA framework can also be applied to design efficient sets of seeds. 
Similar to \cite{Sun2005}, we can successively find seeds that locally maximize the conditional sensitivity, given that the seeds already present in the set do not hit an alignment.
Such a conditional sensitivity can be computed by making the existing seeds' accepting states absorbing without counting a match, so only instances of the current seed candidate are counted.
Evaluating each seed candidate and picking the best one yields the seed to be added to the set.


\section{Protein Fragment Mass Statistics}
\label{sec:fragment_masses}
Peptide mass fingerprinting (PMF) is a technique for protein identification based on mass spectrometry (MS) and database search. Here, we present a PAA to compute the significance of protein identification by PMF.
The protein of interest is enzymatically cleaved into smaller peptides, whose masses are determined by MS. 
Masses are measured in Dalton (Da) or unified atomic mass units (u), where $\unit[1]{Da} = \unit[1]{u}$ equals one twelfth of the mass of an isolated atom of carbon-12 (${}^{12}$C) at rest and in its ground state, or \unit[$1.66053878 \cdot 10^{-24}$]{g}.
The set of peptide masses, the so-called peptide mass fingerprint, is then used to query a database of known proteins. Each database sequence is processed \textit{in~silico} to obtain a theoretical fingerprint, and experimental and theoretical fingerprints are subsequently compared and scored.
The highest-scoring database protein is the most likely candidate for the unknown sample.

In order to come up with a reasonable scoring for such a comparison, one is interested in the probability that a certain peptide fragment mass occurs by chance. 
For example, a mass of \unit[445.2]{Da} is characteristic for the short fragment \texttt{DVCK} (aspartatic acid, valine, cysteine, lysine), which occurs in many known proteins, and is therefore not a helpful feature for protein identification.

Additionally, fragment masses vary for several reasons:
The constituting atoms have different isotope masses; about 98.9\% of all carbon atoms, for example, have six neutrons while about 1.1\% have seven neutrons and are therefore heavier.
Post-translational modifications (additions of chemical groups to some amino acids of the protein) may occur.
Some cleavage sites may be missed by the protease, and one may observe one larger mass that corresponds to the sum of two expected masses.

Therefore, we are interested in the mass distribution of proteolytic fragments, or the joint length-mass distribution of such fragments, to subsequently answer questions about the probability that at least one fragment with a given approximate mass exists in a protein of given or typical length.

An approach chosen by many PMF software packages is to avoid probability computations and use empirical frequencies of fragments in large protein databases instead.
Kaltenbach~\cite{Kaltenbach2007} introduced p-value-based scores that are database-independent,
but based on a text model that represents protein sequences as \iid random strings, \ie as sequences of independent random variables that take values from the alphabet of amino acids according to frequencies estimated from the Swissprot database~\cite{Bairoch1991,Boeckmann2003}.

Here, we generalize this result and present a PAA based on arbitrary finite-memory text models. 
First, we construct a DAA encoding the enzymatic cleavage reaction. Second, from this DAA and a model for random sequences of amino acids, a PAA is constructed by means of Lemma~\ref{lem:daapaa}. 
The resulting PAA provides statistics of peptides usually measured in an MS experiment, in particular, the length distribution and the joint length-mass distribution of such fragments.
Furthermore, we compute the probability that the peptide mass fingerprint of a random protein contains at least one peptide of a certain mass.
This, in turn, provides a significance value for PMF identifications and can be used to derive a scoring scheme, as explored in~\cite{Kaltenbach2007}.
We further incorporate the influence of isotopic distributions, incomplete cleavage, and post-translational modifications by appropriate modifications of the PAA.

\subsection{PAA for Statistics of Proteolytic Fragments}
\label{subsec:paacons}
Cleavage enzymes cut proteins at well-defined places which are often determined by one or two adjacent amino acids (see~\cite{Wang2003,Kaltenbach2007}).
For instance, the most widely used cleavage agent Trypsin cleaves after lysine (K) and arginine (R) unless the next amino acid is proline (P). Such cleavage sites can be described by $\Gamma\bar{\Pi}$ with \emph{cleavage characters} $\Gamma\subset\Sigma$ and \emph{prohibition characters} $\Pi\subset\Sigma$, where $\bar{\Pi}$ denotes the complement of $\Pi$ in the amino acid alphabet
$\Sigma=\{\text{A},\dots, \text{Z}\}\setminus \{\text{B,J,O,U,X,Z}\}$.

Due to their distinct structure, we distinguish the first fragment $F_{1}$ from following fragments $F_{+}$. While the first fragment may start with a prohibition character, following fragments do not. For \iid text models, statistics for all following fragments are \iid as well (assuming an infinite string length), see~\cite{Wang2003,Kaltenbach2007}. For arbitrary finite-memory text models, however, statistics of fragments $F_2,F_3,\ldots$ may differ as well. In any case, statistics for fragments $F_k$ converge as $k$ grows to infinity.
We write $\fragmentX$ to denote an arbitrary fragment (either first or following).
The last fragment does not necessarily end with a cleavage character because of the finiteness of protein sequences.

First, we construct a DAA $\daa=\big(\stateset, \state_0, \Sigma, \delta, \valset, \val_0, \emiset, (\daaemi_\state)_{\state\in \stateset}, (\op_\state)_{\state\in \stateset} \big)$ that sums up amino acid masses (for now, we ignore isotopes) until a cleavage point is encountered.
We set $\stateset:=\Sigma\cup\{\state_0,\bullet, \circ\}$, where $\state_0$ is the start state.
For each state $\state\in\Sigma$, the emission $\daaemi_\state$ gives the mass of amino acid $q$; the set of possible values is $\valset:=\R^+$, where $v_0:=0$ is the start value.
Note that, despite the infinity of $\valset$, the number of values reachable in $\totalsteps$ steps ($\valsize_\totalsteps$) is finite for all $\totalsteps$ (see Section~\ref{sec:statevalue}).
For states $\state'\in\{\state_0,\bullet, \circ\}$ no mass is emitted, that means, $\daaemi_{\state'}=0$.
To sum amino acid masses, we define operations by $\op_\state:(\val,\emi)\mapsto\val+\emi$ for all $\state\in\stateset$.
The transition function $\delta:\stateset\times\Sigma\to\stateset$ reflects the cleavage rules:
\[
 \delta(\state,\sigma)=
\begin{cases}
 \sigma & \text{if } q\in(\Sigma\setminus\Gamma)\cup\{\state_0\}\,,\\
 \sigma & \text{if } q\in\Gamma, \sigma\in\Pi\,,\\
 \bullet & \text{if } q\in\Gamma, \sigma\notin\Pi\,,\\
 \circ & \text{if } q\in\{\bullet, \circ\}\,.
\end{cases}
\]
As can be easily verified, this DAA has the property that $\delta(\state_0,\prefix{s}{i+1})=\bullet$ if and only if the first fragment in $s$ has length $i$. Then, $\daaval_{\daa}(\prefix{s}{i+1})=m$, where $m$ is the mass of the first fragment.

\begin{figure}[tb]\centering
	\includegraphics[width=0.90\textwidth]{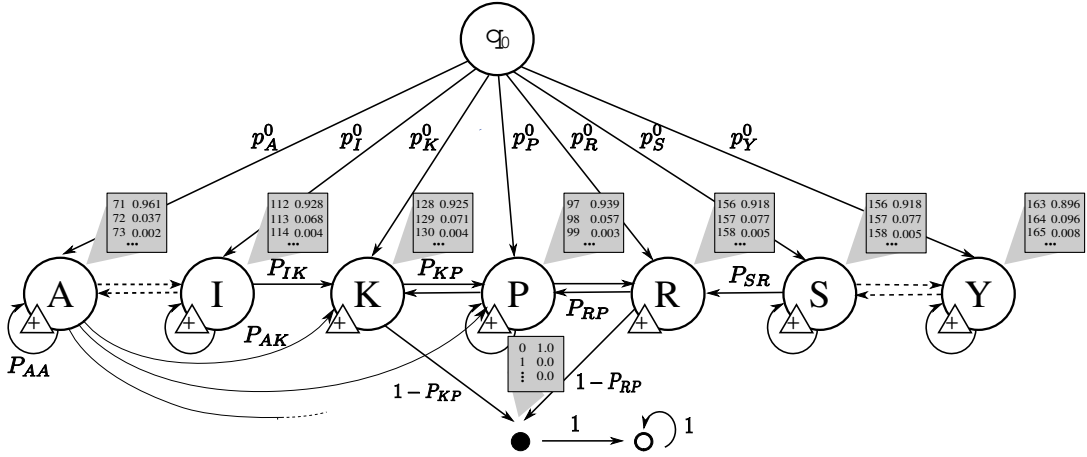}
	\caption{Sketch of the PAA measuring the mass of the first fragment resulting from tryptic cleavage of a random protein. Following fragments are handled by a modified PAA where the initial distribution is normalized by $1-p_{P}$ and start in $P$ is prohibited since following fragments cannot start with a prohibition character. For the sake of simplicity, not all transitions are shown. The states' weight distributions correspond to the respective isotopic distributions. Here, integer masses are displayed. The operation associated to each state is ``$+$''.}
	\label{fig:pmfpaa}
\end{figure}

From the DAA and a text model, a PAA is obtained by using Lemma~\ref{lem:daapaa}. 
To take the isotopic mass distribution of each amino acid into account, each state's emission distribution can be changed accordingly. 
For practical computations, it is sufficient to perform calculations up to a limited mass accuracy as the accuracy of MS instruments is limited as well. 
One could, for example, consider masses up to one decimal position.
A sketch of a PAA for first fragments and a first-order Markovian text model is shown in Figure~\ref{fig:pmfpaa} (including isotopic distributions).
From now on, we assume that masses have been scaled and then rounded to integers to achieve the desired precision.

The difference between the first fragment $F_1$ and the following fragments $F_2, F_3,\ldots$ lies solely in the initial state distribution.
The transition probabilities from the start state to other states need to be adjusted in such a way that (1)~prohibition characters are forbidden at the beginning of a following fragment and (2)~the transitions reflect the distribution of text model states at the end of the previous fragment.
This distribution can be obtained from the PAA for the previous fragment.
Note again that statistics for all fragments $F_k$ with $k>1$ are \iid if the underlying text model is \iid as well.

\subsection{Mass and Length Distributions}
\label{subsec:peptidestat}
By means of the constructed PAAs, we compute fragment statistics as the length distribution and the joint length-mass distribution.
We denote the length of a fragment by $L(\fragmentX)$ and the (integer) mass of a fragment by $M(\fragmentX)$.
The length of a random fragment generated by the PAA corresponds to the waiting time for reaching state~$\bullet$ minus one (as the last processed character does not belong to the fragment).
When only the distribution of fragment lengths (and not the mass distribution) is required, we can use Lemma~\ref{lem:waiting_state} to compute the distribution of this waiting time.

Generally, the joint state-value distribution $f_{\totalsteps}(\state,\val) = \prob(\stateproc_{\totalsteps}=\state,\valproc_{\totalsteps}=\val)$ discussed in Section~\ref{sec:statevalue} yields the joint length-mass distribution~$\fragmentXdist$ of $\fragmentX$:
\[
     \fragmentXdist(\totalsteps,m):= \prob\big(L(\fragmentX)\!=\!\totalsteps,\, M(\fragmentX)\!=\!m\big)
   = f_{\totalsteps+1}(\bullet,m) = \prob(\stateproc_{\totalsteps+1}\!=\!\bullet,\valproc_{\totalsteps+1}\!=\!m)\,.
\]
The computation of $f_{\totalsteps+1}$ takes $\OO(\totalsteps \cdot |\stateset|^2 \cdot \valsize_{\totalsteps} \cdot |\emiset|)$ time and $\OO(|\stateset| \cdot \valsize_\totalsteps)$ space (see Lemma~\ref{lem:paa_basic}).
Since both $|\stateset|$ and $|\emiset|$ are constants, and the range of possible masses $\valsize_\totalsteps$ grows linearly with both $\totalsteps$ and the mass scaling factor (or mass precision) $\lambda$, we need $\OO(\lambda\totalsteps^2)$ time and $\OO(\lambda\totalsteps)$ space.

\subsection{Mass Occurrence Probabilities}
For the interpretation of mass spectra, the most important quantity is the mass occurrence probability of a  measured mass~$m$, \ie the probability that the fragmentation of a random protein sequence contains at least one fragment of mass~$m$.
To account for possibly inaccurate measurements, it is advisable to compute the probability that a fragment in a certain mass range $[m-\Delta,m+\Delta]$ occurs.
We modify the PAA such that it does not move into the absorbing state~$\circ$ once the first fragment has ended.
We therefore remove state~$\circ$ and replace~$\bullet$ by $|\Sigma|$ states named $\bullet_\sigma$ for each $\sigma\in\Sigma$.
That means, $\stateset:=\{\state_0\}\cup\Sigma\cup\{\bullet_\sigma\,:\,\sigma\in\Sigma\}$.
Being in state $\bullet_\sigma$ means that a fragment has ended and the first character of the next fragment is $\sigma$.
Therefore, the emission distribution of $\bullet_\sigma$ is the same as for state $\sigma$.
In terms of the transition function $\delta$, each state $\bullet_\sigma$ acts like state $\sigma$.
Formally,
\[
\delta(\state,\sigma):=
\begin{cases}
 \sigma & \text{if } q\in(\Sigma\setminus\Gamma)\cup\{\state_0\}\,,\\
 \sigma & \text{if } q\in\Gamma, \sigma\in\Pi\,,\\
 \bullet_\sigma & \text{if } q\in\Gamma, \sigma\notin\Pi\,,\\
 \delta(\sigma',\sigma) & \text{if } q=\bullet_{\sigma'}\,.
\end{cases}
\]
To keep track of whether a fragment in the given mass range has already been observed, we introduce an absorbing value $\diamond$; that means, all operations are adapted such that once the value $\diamond$ has been attained, it is not changed.
The new states $\bullet_\sigma$ get the following operations:
\[\op_{\bullet_\sigma}(\val,\emi) :=
\begin{cases}
\diamond & \mbox{if }\val\in[m-\Delta,m+\Delta]\,, \\
e & \mbox{otherwise}\,.
\end{cases}
\]
The probability that a protein sequence of length~$\totalsteps$ contains a segment in the mass range $[m-\Delta,m+\Delta]$ is given by
$\prob(V_\totalsteps=\diamond) + \prob\big(V_\totalsteps\in [m-\Delta,m+\Delta]\big)$,
where the second summand accounts for the probability that the last fragment has the sought mass.

\subsection{Missed Cleavages and Post-Translational Modifications}
\label{subsec:ptm}
Protein identification by MS is complicated by the occurrence of partial enzymatic cleavage which results in peptides with internal missed cleavage sites. 
Even for Trypsin, which is reported to have a high cleavage specificity, incomplete digestion is not uncommon~\cite{Siepen2007,Matthiesen2007}.
The presented PAA can be modified to account for missed cleavages by adjusting the transition probabilities outgoing from the cleavage characters:
The probability to end a fragment after reading a cleavage character is reduced, while transitions from a cleavage character to non-prohibition characters occur with a small probability. 
Moreover, in a Markovian text model, one can include information about missed cleavage patterns, \ie the probabilities can be weighted according to the amino acid propensities in proximity to missed cleavage sites. 
 
Another complicating issue is that many proteins are modified after translation. 
A \emph{post-translational modification} (PTM) is a chemical process that changes the properties of a protein. 
It includes the addition of functional groups such as acetate or phosphate, the modification of amino acids, or structural changes such as the formation of disulfide bridges. 
Putative modifications can themselves be discovered by means of MS, comparing the mass measured to the mass expected for the identified protein or peptide \cite{Annan1997,Mann2003,Tsur2005}.
PTMs result in a change in the molecular mass of the protein. 
Examples of common modifications include phosphorylation (+\unit[80]{Da}), acetylation (+\unit[42]{Da}), and methylation (+\unit[14]{Da} or +\unit[28]{Da}). 
An overview of important PTMs along with the resulting mass shifts in \unit[]{Da} is provided in the review of Mann and Jensen~\cite{Mann2003}.
The PAA presented above can be modified to incorporate PTMs as follows:
``Global'' modifications of the protein are incorporated into the mass distribution of either the start state $\state_0$ or of an end state $\circ$ or $\bullet$. 
Amino acid-specific modifications are incorporated by modifying the mass distribution of that particular amino acid state, possibly splitting the state into several copies accounting for different modification probabilities. Reasonable frequencies of PTMs associated with particular amino acids are provided in a study of Tsur~et~al.\ \cite{Tsur2005}.

\subsection{Characteristic Masses of Protein Families}
PAAs provide the possibility to combine any probabilistic protein model with a cleavage scheme and, hence, to obtain a probabilistic PMF for the model.
This allows to compute statistics of ``random'' peptide fragments, as shown above, but also to compute characteristic masses of protein families, which are often represented by probabilistic models, such as HMMs in the Pfam database~\cite{FinnMistryTate+2010Pfam}. As character-emitting HMMs are special cases of finite-memory text models (see Section~\ref{sec:text_model}), we can combine a Pfam model with a DAA to compute fragment statistics specific to the protein family.

Mass~$m$ is said to be specific for a protein family (for a particular cleavage scheme, with respect to a set of protein families) if the occurrence probability of fragments with mass~$m$ is greater than $1-\varepsilon$ for some small~$\varepsilon>0$
and if fragments of mass~$m$ are not expected in any other protein family.
We are not aware of any work done in this area; hence it is unkown if family-specific fragment masses exist for any cleavage scheme.

\section{High-Throughput Sequencing}
\label{sec:sequencing}

With the 454 sequencing technology, up to a million DNA reads of lengths up to 400 nucleotides can be determined in parallel by pyrosequencing.
Nucleotides are successively synthesized to single-stranded DNA templates evoking an enzymatic cascade, resulting in detectable flashes of light.
These signals are reported in a so-called \emph{pyrogram} from which the sequence of nucleotides (the \emph{read}) is determined.
In 454 sequencing, the four different nucleotides are sequentially flowed over the reaction plate which holds the DNA templates.
The particular order of nucleotide flows is called \emph{dispensation order} and repeated for several (say, 100) cycles.
The 454 systems GS20 and GS-FLX use the standard dispensation order TACG.
Whenever the next nucleotide in the template sequence matches the currently dispensed nucleotide (in reality, its complement, but this is a technical detail), the dispensed nucleotide extends the complementary strand of the template.
In case of a homopolymer run (the same type of nucleotide appearing several times consecutively), the strand is extended by the corresponding number of nucleotides at once.
Such an event is detected by a correspondingly more intense flash of light.
The more recent IonTorrent technology has the same characteristics, but uses a different underlying technology (change in pH instead of light detection).
Table~\ref{tab:disporder} shows how the sequence of a template fragment is determined with two different dispensation orders.

\begin{table}[t]
\caption{Depending on the dispensation order, different numbers of nucleotides can be recognized during three cycles. The dispensation orders \texttt{TACG} and \texttt{GTCA} are compared for the sequence \texttt{GTCGTATCCC}.
Note the homopolymer run of three \texttt{C}s, which is sequenced with a single flow.}
\label{tab:disporder}
\begin{center}
\begin{tabular}{cccp{5mm}cc}
\hline\hline
      & \multicolumn{2}{c}{Dispensation order \texttt{TACG}} & & \multicolumn{2}{c}{Dispensation order \texttt{GTCA}}\\\hline
Cycle & Nucleotide     & Recognized? & & Nucleotide     & Recognized?\\\hline
1 & \texttt{T} & -       & & \texttt{G} & $\surd$ \\
  & \texttt{A} & -       & & \texttt{T} & $\surd$ \\
  & \texttt{C} & -       & & \texttt{C} & $\surd$ \\
  & \texttt{G} & $\surd$ & & \texttt{A} & - \\
2 & \texttt{T} & $\surd$ & & \texttt{G} & $\surd$ \\
  & \texttt{A} & -       & & \texttt{T} & $\surd$ \\
  & \texttt{C} & $\surd$ & & \texttt{C} & - \\
  & \texttt{G} & $\surd$ & & \texttt{A} & $\surd$ \\
3 & \texttt{T} & $\surd$ & & \texttt{G} & - \\
  & \texttt{A} & $\surd$ & & \texttt{T} & $\surd$ \\
  & \texttt{C} & -       & & \texttt{C} & \phantom{$\surd\surd$}$\surd\surd\surd$\\
  & \texttt{G} & -       & & \texttt{A} & -\\\hline
\multicolumn{2}{l}{Read length:} & 6 & & & 10 \\\hline \hline
\end{tabular}
\end{center}
\end{table}

We investigate the exact length distribution of 454 sequencing reads, assuming a fixed number of nucleotide flows, a given dispensation order, and an arbitrary finite-memory text model.
Previously, Rahmann~\cite{Rahmann2006} studied the combinatorics of sequences that can be reliably sequenced by the 454 technology.
Kong \cite{Kong2009a} obtained generating functions for the length distribution, but the dispensation order was restricted to a permutation of the nucleotides, and the tex model was restricted to an \iid model.

Here we design a PAA that allows for arbitrary finite-memory text models and arbitrary dispensation orders: Every ordering of nucleotides, where each nucleotide occurs at least once and no nucleotide is flowed twice consecutively, is reasonable, e.g., TCGACG.

\subsection{PAA for the Length Distribution of 454 Reads}
Instead of attacking the length distribution problem directly, we first construct a DAA that computes the number of nucleotide flows needed to read (a finite prefix of) an input sequence with a given dispensation order $d=d[0]\dots d[\ell-1]$.
In other words, in each step, the DAA reads one nucleotide to be sequenced, and the emitted value models the waiting time (number of flows) since the previous sequenced nucleotide.
We cast the length distribution problem as a waiting time problem by waiting for a fixed number $f$ of flows (Section~\ref{sec:waitingtimes}; waiting for a value, Lemma~\ref{lem:waiting_value}).
As a consequence, we obtain the exact length distribution $\dist(L)=\dist\big(L(f,d)\big)$ of reads sequenced after $f$~nucleotide flows according to a dispensation order $d$.
In practice, $f$ is usually a multiple of $\ell=\len{d}$.

We define a DAA with state set
$\stateset^\daa := \left(\Sigma\times\{0,\dots,\ell -1\}\right)\cup \bigcup_{i=-1}^{\ell -1}\, \{(\epsilon,i)\}$ with start state $\state_0 := (\epsilon,-1)$ and attach the following semantics: 
Being in state $(\sigma,j)$ means that the last two sequenced nucleotides were $\sigma$ and $d[j]$. 

The (deterministic) emission of state $(\sigma,j)$ tells us how many flows it took to reach nucleotide $d[j]$ after $\sigma$:
If $d[j]=\sigma$, they are part of the same homopolymer run and the emission is zero; otherwise, it is the smallest forward distance between these nucleotides in the (cyclically interpreted) dispensation order.
For example, if $d=$ TCGACG, $\sigma=$ G, and $j=1$, so $d[j]=$ C, then the emission is $2$ (skipping over $d[0]=$ T). Formally,
\begin{equation}
 \daaemi_{(\sigma,j)}:=\min \big\{ i\in\{0,\ldots,\ell-1\}\ \big|\ d[j-i\mod\ell]=\sigma \big\}.
\end{equation}
The transition target from state $(\sigma,j)$ when reading character $\sigma'$ is consequently
\[    \delta\big((\sigma,j),\sigma'\big)
   := \Big(d[j]\,,\; j+\min\big\{i\in\{0,\ldots,\ell-1\}\,\big|\,d[j+i\mod\ell]=\sigma'\big\}\!\mod\ell\Big);
\]
this is also valid for $\sigma=\epsilon$.
Intuituvely, the ``old'' $j$ becomes the ``new'' $\sigma=d[j]$, and the ``new'' $j$ is obtained by cyclically searching forward in the dispensation order.
The initial transition targets and emissions are given by
\begin{align*}
 \delta\big((\epsilon,-1),\sigma'\big)
   &:= \big(\epsilon\,,\;\min\big\{i\in\{0,\ldots,\ell-1\}\,\big|\,d[i]=\sigma'\big\}\big); \\
 \daaemi_{(\epsilon,j)} &:= j+1 \quad \text{for } j\in\{-1,\dots,\ell-1\}.
\end{align*}
The sequencing protocol specifies the number $f$ of nucleotide flows; typically $f=400$ for the standard dispensation order TACG, \ie 100 cycles.
The operation in each state adds the emitted number of flows, truncating at $f+1$.
This completes the construction of a DAA.

We are interested in the sequence length distribution up to a given length~$\tmax$.
We define the target set $\mathcal{T}:=\{f+1\}$ (cf.\ Definition~\ref{def:waitingtime})
and obtain the read length as one less than the waiting time
\[ W_{\mathcal{T}} = \min \left\{ t\in\N_0 \,\middle|\, V_{t} = f+1 \right\}. \]
Invoking Lemma~\ref{lem:daapaa} yields the corresponding PAA for a general finite-memory text model, from which we obtain the read length distribution up to a given~$n$.
Applying Lemma~\ref{lem:daapaa:time}'s statement about waiting times with $\Sigma=\OO(1)$, $|\mathcal{C}|=\OO(1)$, $|\stateset^\daa|=\OO(\ell)$ and $\valsize_{\totalsteps} = \OO(\ell \totalsteps)$ results in a running time of $\OO(\totalsteps^2\, \ell^2)$ and a space requirement of $\OO(\totalsteps \ell^2)$.

As an application, we compared the expected read lengths under an estimated first-order Markov model for all reasonable dispensation orders of length $4$ to $10$ on two datasets of 454 reads of yet unfinished strains of the GC-rich bacteria \textit{S.~meliloti} and \textit{R.~Lupinii}, provided by the Genetics Department of Bielefeld University.
We observed that the standard machine settings could be improved to yield approximately $10\%$ longer reads on average, namely \unit[282]{nt} instead of \unit[257]{nt} with the GS-FLX instrument during $f=400$ nucleotide flows.


\section{Discussion}
\label{sec:discussion}

We have presented the concept of probabilistic arithmetic automata that blends well into the landscape of existing stochastic models like Markov chains and hidden Markov models. 
In fact, PAAs can be seen as Markov chains over a larger state space. 
The benefit of PAAs lies in their utility as a modelling technique, specifically (1) the required state space is often small, (2) the connection between states and values becomes evident, and (3) an elegant notation and well-arranged recurrences are obtained.
As shown in the second part of this article, many applications can conveniently be approached using the PAA framework. 
The method is especially suited for applications involving the deterministic processing of random texts. 
In these cases, a PAA can be constructed from a deterministic arithmetic automaton and a finite-memory text model, a quite general class of text models that covers simple \iid models as well as arbitrary-order Markov models and character-emitting HMMs.

\section*{Acknowledgments}
The authors wish to thank Jens Stoye and Sebastian B\"ocker for their valuable comments.
Timo St\"ocker carefully read the manuscript.
IH was supported by the NRW Graduate School in Bioinformatics and Genome Research, Bielefeld. 

\bibliographystyle{plain}
\bibliography{lit}

\end{document}